\documentclass[fleqn,10pt]{wlscirep}
\usepackage[utf8]{inputenc}
\usepackage[T1]{fontenc}

\usepackage{lineno,hyperref}
\usepackage{amsmath,graphicx,makecell}
\usepackage{enumitem}
\graphicspath{{./fig/}}
\usepackage{threeparttable}
\usepackage{booktabs}
\usepackage{algorithmic}
\usepackage{multirow}
\usepackage{multicol}
\usepackage{tabularx} 
\usepackage[linesnumbered, ruled, vlined]{algorithm2e}
\usepackage{footnote}

\title{A Cross-Domain Approach to Analyzing the Short-Run Impact of COVID-19 on the U.S. \\ Electricity Sector}
\begin{document}
\author[1,2]{Guangchun~Ruan}
\author[1]{Dongqi~Wu}
\author[1]{Xiangtian~Zheng}
\author[2]{Haiwang~Zhong}
\author[2]{Chongqing~Kang}
\author[3]{Munther A. Dahleh}
\author[1,*]{S. Sivaranjani}
\author[1,*,$\dagger$]{Le~Xie}

\affil[1]{Department of Electrical and Computer Engineering, Texas A\&M University, College Station, TX 77843, USA}

\affil[2]{Department of Electrical Engineering, Tsinghua University, Beijing 100084, China}

\affil[3]{Institute for Data, Systems, and Society, Massachusetts Institute of Technology, Cambridge, MA 02139, USA}

\affil[*]{Co-last author.}

\affil[$\dagger$]{Corresponding author: le.xie@tamu.edu}

\begin{abstract}

\vspace{-2em}
The novel coronavirus disease (COVID-19) has rapidly spread around the globe in 2020, with the U.S. becoming the epicenter of COVID-19 cases since late March.  As the U.S. begins to gradually resume economic activity, it is imperative for policymakers and power system operators to take a scientific approach to understanding and predicting the impact on the electricity sector. Here, we release a first-of-its-kind cross-domain open-access data hub, integrating data from across all existing U.S. wholesale electricity markets with COVID-19 case, weather, {mobile device} location, and satellite imaging data. Leveraging cross-domain insights from public health and mobility data, we rigorously uncover a significant reduction in electricity consumption that is strongly correlated with the number of COVID-19 cases, degree of social distancing, and level of commercial activity.
\vspace{-1.8em}
\end{abstract}

\flushbottom
\maketitle
\thispagestyle{empty}

\section*{Introduction}
As the U.S. {responds to the} novel coronavirus disease (COVID-19) and states re-open the economy, there is much uncertainty regarding the duration and severity of the impact on the electricity sector. Given the rapid spread of COVID-19 and the corresponding policy changes, there has been relatively little scholarly work on the impact of COVID-19 on the electricity sector. Several reports from both {peer-reviewed\cite{gillingham2020short,le2020temporary} and non-peer-reviewed venues such as news media\cite{nyt}, social media\cite{RN25,RN24,RN26}, consulting firms\cite{brattle,wood-mck}, non-profit organizations\cite{RN16}, government agencies\cite{RN22,RN27},} and professional communities\cite{skoeei,SGN}, have shed some light on the adverse impact  on the electricity {and clean energy} sectors, including operational reliability degradation, decrease in wholesale prices, and delayed investment activities. Electricity consumption analyses from regional transmission organizations (RTOs)\cite{RN29,RN30,RN31} also suggest an overall reduction in energy consumption, especially in zones with large commercial activity.

However, such assessments are still at a nascent stage, with several gaps in existing research. First, the lack of consistent assessment criteria renders results across distinct geographical locations incomparable. Second, { several} existing statistical analyses do not rigorously calibrate a baseline electricity consumption profile in the absence of the pandemic considering the influence of exogenous factors like the weather. Finally, cross-domain data like public health data (COVID-19 cases and deaths) and social distancing data ({mobile device} location) that can provide valuable insights have not been considered so far in the analysis of the electricity sector. 

Here, we develop a cross-domain open-access data hub, COVID-EMDA$^+$ (Coronavirus Disease and Electricity Market Data Aggregation$^+$), to track and measure the impact of COVID-19 on the U.S. electricity sector\cite{RN1}. This data hub integrates information from electricity markets with heterogeneous data sources like  COVID-19 public health data, weather, {mobile device} location information and satellite imagery data, that are typically unexplored in the context of the energy system analysis. The integration of these cross-domain data sets allows us to develop a novel statistical model that calibrates the electricity consumption based on mobility and public health data, which have otherwise not been considered in conventional power system load analysis literature thus far.  Leveraging this cross-domain data hub, we 
uncover and quantify a ``delayed'' impact of the number of COVID-19 cases, social distancing, and mobility in the retail sector on electricity consumption. In particular, the diverse time-scales and magnitudes of top-down (federal or state policies and orders) and bottom-up (individual-level behavior change in social distancing) responses to the pandemic collectively influence the electricity consumption in a region. 
We observe a significant reduction in electricity consumption across all U.S. markets {(ranging from $6.36\%$ to $10.24\%$ in April, and $4.44\%$ to $10.71\%$ in May)}, which is strongly correlated with the rise in the number of COVID-19 cases, the size of the stay-at-home population (social distancing), and mobility in the retail sector (representative of the share of commercial electricity use), which emerges as the most significant and robust influencing factor.  

\section*{Cross-domain Data Hub: COVID-EMDA$^+$}
We first develop a comprehensive cross-domain open-access data hub, COVID-EMDA$^+$ (Coronavirus Disease and Electricity Market Data Aggregation$^+$)\footnote{The $+$ symbol in COVID-EMDA$^+$ indicates the integration of cross-domain data sets like public health and mobility data with conventional electricity market data.}, publicly available on Github~\cite{RN1}, integrating electricity market, weather, {mobile device} location, and satellite imaging data into a single ready-to-use format. The original sources for each dataset are detailed in the Data and Code Availability section. We pay special attention to the impact of COVID-19 on electricity markets in the U.S.~\cite{RN32} for two reasons. First, electricity market data are usually timely, accurate, abundant and publicly available in the U.S., making the market dataset ideal for impact tracking and measurement. Second, wholesale electricity markets in U.S. cover the top eight hardest-hit states, and more than $85\%$ of the national number of confirmed COVID-19 cases as of May 2020 (Supplementary Fig. S-1-c). 

There are seven regional transmission organizations (RTOs) or electricity markets in the U.S., namely, California (CAISO)~\cite{RN2}, Midcontinent (MISO)~\cite{RN3}, New England (ISO-NE)~\cite{RN4}, New York (NYISO)~\cite{RN5}, Pennsylvania-New Jersey-Maryland Interconnection (PJM)~\cite{RN6}, Southwest Power Pool (SPP)~\cite{RN7}, and Electricity Reliability Council of Texas (ERCOT)~\cite{RN8}. For each regional market, we aggregate data pertaining to the load, generation mix, and day-ahead locational marginal price (LMP). To improve the overall data quality, we also integrate market data from the Energy Information Administration (EIA)~\cite{RN9} and EnergyOnline company~\cite{RN10}. The major challenges in integrating raw electricity market data into a unified framework are summarized in the Methods section. We integrate the electricity market data with weather data\cite{RN12} (temperature, relative humidity, wind speed and dew temperature)  from the National Oceanic and Atmosphere Administration (NOAA). We will use this data to estimate an accurate baseline electricity consumption profile taking into account weather, calendar, and economic factors (annual GDP growth rate), against which the impact of COVID-19 will be quantified. 
\begin{figure}[!b]
	\centering
	\includegraphics[width=0.9\textwidth]{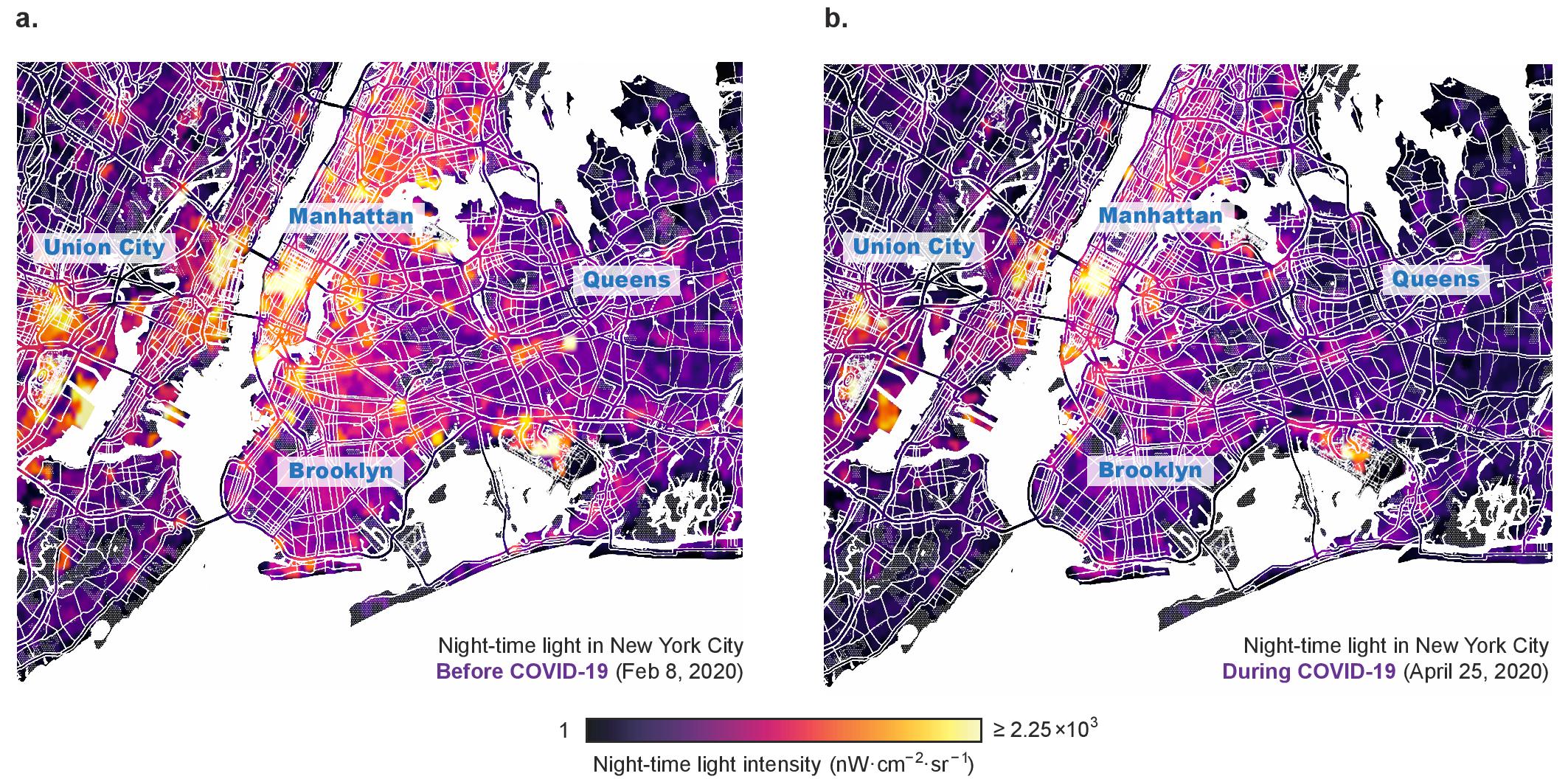}
	\caption{Visualization of the impact of COVID-19 on electricity consumption using NTL data for New York City:  (a) NTL imagery before the outbreak of COVID-19 (February 8, 2020). (b) NTL imagery during the outbreak (April 25, 2020). The sampling time of both representative snapshots is $1$ a.m. on Saturday, when the sky is clear of cloud. The raw data are pre-processed to filter out ambient noise and focus on only the urban area of the city. A colormap is used to clearly illustrate the light intensity, in which bright color indicates strong light and dark color indicates dim light. The background city map is retrieved from OpenStreetMap\cite{OSM}.} 
	\label{fig-ntl}
\end{figure}

To obtain further cross-domain insights,  we integrate public health data on COVID-19 cases from {multiple sources}~\cite{RN17,xu2020epidemiological,yang2020covidnet} and {mobile device} location data from SafeGraph~\cite{sg_social,sg_pattern} comprising of  county-level social distancing data and pattern of visits to Points of Interest (POIs) like restaurants and grocery stores (see Supplementary Note SN-2 for a detailed description). We aggregate the {mobile device} location data by county and POI category, and define the \textbf{\textit{stay-at-home population}} and the \textbf{\textit{population of on-site workers}} (indicative of the social distancing level) as the estimated number of people who stay at home all day, and the number of people who work at a location other than their home for more than 6 hours on a typical working day, respectively. The \textbf{\textit{mobility in the retail sector}}, defined as the number of {{visits to retail establishments}} per day (see Supplementary Note SN-3 for a list of 25 included merchant types) is also of interest, since it is indicative of the level of commercial activity. 
Finally, we  integrate satellite imagery from the NASA VNP46A1 "Black-Marble" \cite{BlackMarble} dataset into the COVID-EMDA$^+$ hub as a tool for visualizing the impact of COVID-19 on electricity consumption (see Supplementary Note SN-1 for a detailed description of this dataset). The complete architecture of the data hub is shown in Supplementary Fig. S-1-a. { The detailed description of all the original data sources, and a summary of the utility of each cross-domain data source, are provided in Supplementary Note SN-4.}

Using night-time light (NTL) data from satellite imagery, Fig. \ref{fig-ntl} visualizes the impact of COVID-19 on electricity consumption for New York City (see Supplementary Note SN-1 for a detailed description of how the NTL data is processed to obtain these plots). 
The reduction in NTL brightness provides a strong visual representation of the effect of COVID-19 on electricity consumption level in such major {urban areas} (see Supplementary Fig. S-2 for NTL visualization of other metropolises), { where a significant component of the electricity consumption comprises of large commercial loads. This result serves as a preview of the insights that emerge from the statistical analysis in the following sections, namely, that the level of commercial activity (quantified by mobility in the retail sector in our later analysis) is a key contributing factor for the change in electricity consumption during COVID-19.} In the following analysis, we will leverage {the} cross-domain {COVID-EMDA$^+$} data hub to  quantify this reduction of electricity consumption, and demonstrate its correlation with the number of COVID-19 cases, degree of social distancing, and level of commercial activity. 

\section*{Quantifying Changes in Electricity Consumption Across RTOs and Cities in the U.S.}
Following the idea of predictive inference~\cite{rubin2003basic}, we leverage the cross-domain COVID-EMDA$^+$ data hub to derive statistically robust results on the changes in electricity consumption correlated with the COVID-19 pandemic. We achieve this by carefully designing an ensemble backcast model to accurately estimate electricity consumption in the absence of COVID-19, which are then used as benchmarks against which the impact of COVID-19 is quantified. 

We begin by analyzing the reduction in electricity consumption in the New York area, which is the epicenter of the pandemic in the U.S. Fig.~\ref{fig-backcast}-a shows the comparison between actual electricity consumption profile, {ensemble} backcast results (with $10\%-90\%$ and $25\%-75\%$ quantiles), and the electricity consumption profile in previous year (aligned by day of the week using NYISO data; for example, February 4, 2019 and February 3, 2020 are compared because they are both Mondays of the fifth week in the respective year). The strong match between the curve shapes indicates that the {ensemble} backcast estimations reliably verify the insignificant change in electricity consumption before the COVID-19 outbreak (February 3 and March 2) and much larger change afterwards (April 6 and 27). Note that the electricity consumption profile in 2019, although being a common and simple choice in many analyses, is typically an inaccurate baseline for  impact assessment in 2020.

A cross-market comparison, with both the point- and interval-estimation results, is conducted in Fig.~\ref{fig-backcast}-b to show the impact of COVID-19 on different marketplaces. The interval estimation is calculated using the $10\%$ and $90\%$ quantiles which can be regarded as reliable estimation boundaries. The ensemble backcast models successfully capture the dynamics of changes in electricity consumption and provide a reliable statistical comparison among different regions. It is clearly seen that all the markets experienced a reduction in electricity consumption in {both April and May}; however, the magnitudes of the reductions were diverse, varying from $6.36\%$ to $10.24\%$ in April, and {$4.44\%$ to $10.71\%$ in May}.  Additionally, our estimation results for April match well with  official reports~\cite{RN29, RN30, RN31}. According to Fig.~\ref{fig-backcast}-b, NYISO and MISO experienced the most severe reduction in electricity consumption {in both April and May}, while ERCOT and SPP suffered the least. 
{ All electricity markets showed a rebound in electricity consumption in June that may be correlated with partial reopening of the economy; however, the magnitudes of the rebound were once again diverse across markets as seen in Fig.~\ref{fig-backcast}-b.} {Finally,} in dense urban areas, the impact {of COVID-19} was more pronounced, with New York City and Boston experiencing a $14.10\%$ and $11.32\%$ reduction in electricity consumption respectively in April, likely due to the high population density and large share of commercial energy use in these areas. (The same factors explain why Houston, which is more geographically dispersed, was not significantly impacted). We will examine such potentially relevant factors more closely in the following section.

\section*{Impact of Public Health, Social Distancing, and Commercial Activity on Electricity Consumption During COVID-19}
In order to interpret the changes in electricity consumption during COVID-19, we begin by investigating three potential influencing factors, namely, public health (indicated by the number of COVID-19 cases), the social distancing (indicated by the size of the stay-at-home population and the population of on-site workers), and the level of commercial activity (indicated by a reduction in visits to retail establishments). These influencing factors possess two important features that must be taken into account while interpreting their influence on electricity consumption. 

\begin{figure}
    \centering
    \includegraphics[width=0.95\textwidth]{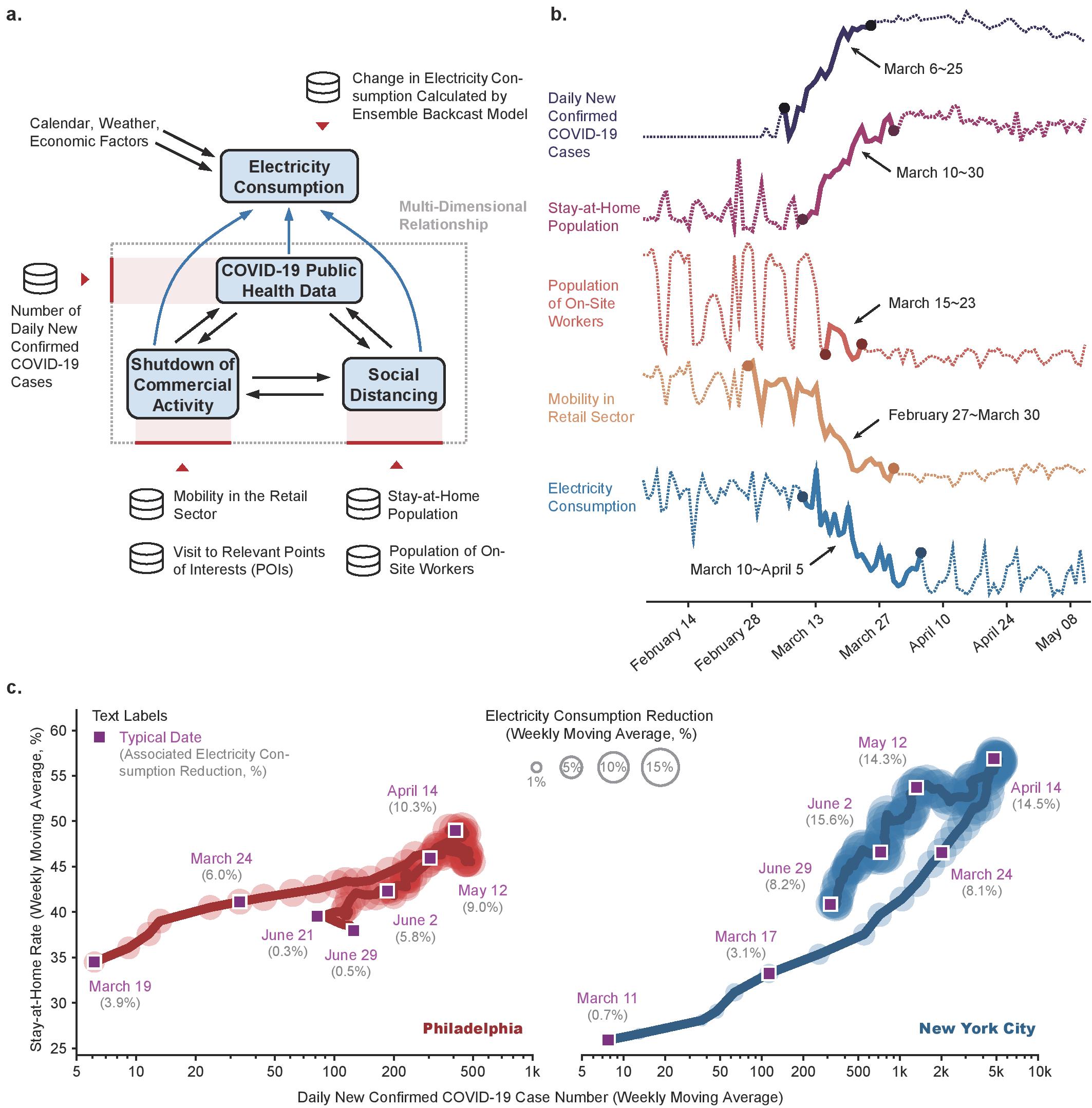}
    \caption{(a) Multi-dimensional relationship between case load, social distancing, shut down of commercial activity and electricity consumption. 
	Heterogeneous data sources from COVID-EMDA$^+$ are applied as indicators of these factors.
	(b) Wide variation in the time scales of different factors influencing electricity consumption during the COVID-19 pandemic. The raw number of confirmed COVID-19 cases are offset by 1 and plotted on a logarithmic scale. {The segments in bold indicate the transition periods for each variable (see Supplementary Fig. S-10 for the details on how these transition periods are defined and identified.).} It is apparent that the electricity consumption started dropping almost immediately after the national emergency declaration. The number of new confirmed cases started to rise significantly a couple days earlier. The stay-at-home population and population of {on-site} workers started changing around the time of the national emergency declaration, while the slight rebound around April 20 coincided with re-opening policies in a few states. The mobility in the retail sector started dropping at the very early stages of the COVID-19 outbreak, due to individual consumer responses to the pandemic. 
	(c) Trace of the evolution of daily new confirmed cases and social distancing, and the associated rate of reduction in electricity consumption for two representative metropolises - New York City and Philadelphia. The bubble sizes indicate the percentage reduction in electrical consumption (with larger bubble sizes indicating more reduction in consumption).  
	The number of COVID-19 cases and the size of the stay-at-home population are smoothed by a weekly moving average to properly extract the trends. Both cities follow a fast developing period in March 2020 and a more stable period afterwards. {A slight rebound in the electricity consumption is also observed in the trace during June 2020.}
	}
    \label{fig-dyn-fac-sd}
\end{figure}

First, there is a complex multi-dimensional relationship between the number of COVID-19 cases, social distancing, shut down rate of commercial activity, and electricity consumption, as shown in Fig.~\ref{fig-dyn-fac-sd}-a. For example, stricter social distancing
\noindent \begin{minipage}{0.99\textwidth}
\vspace{-1em}
   \textbf{a.} 	\\
   \includegraphics[width=\textwidth]{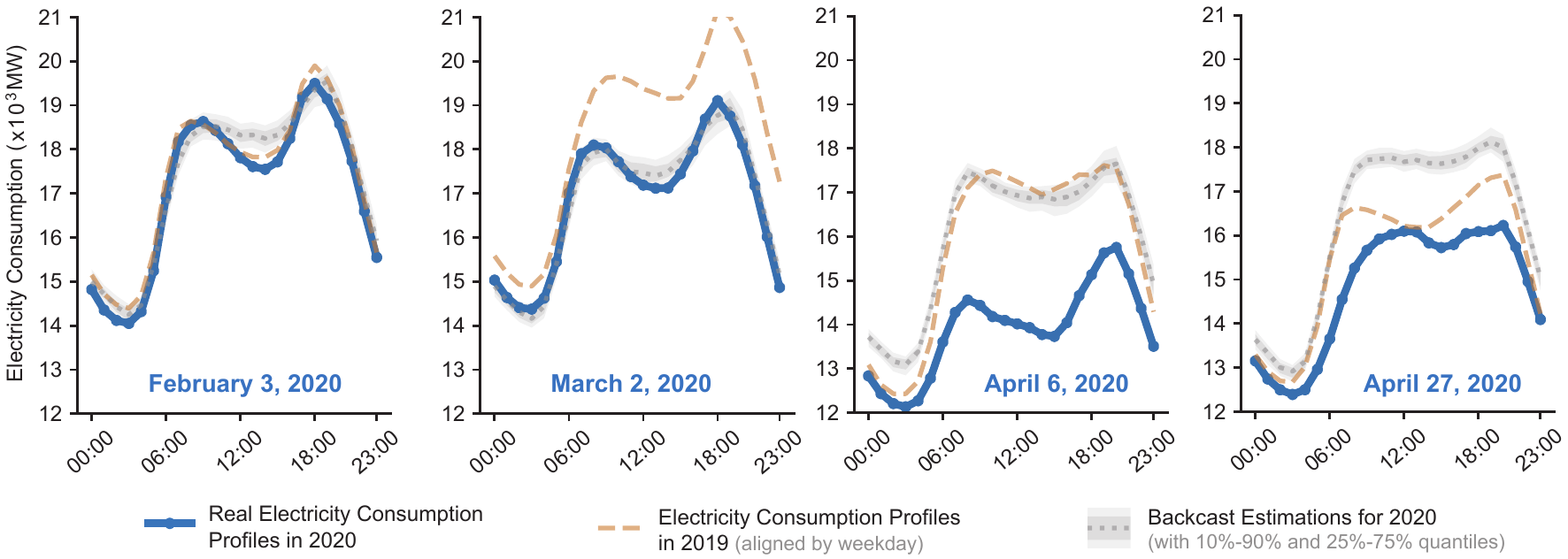}\\
   \vspace{0.4em}
     \textbf{b.} 	\\
    		\label{tab-cross-market-compare}
	\linespread{1.3}\selectfont  
	\setlength\tabcolsep{3pt}  
	\resizebox{\textwidth}{!}{%
    		\begin{threeparttable}
		\begin{tabular}{lclclclclclclc}
		\toprule
			Electricity Consumption Reduction (\%) & CAISO && MISO && ISO-NE && NYISO && PJM && SPP && ERCOT  \\
			\midrule
			Average in February      & $\bf{-1.31}$      					 && $\bf{-0.14}$      				 && $\bf{2.15}$      				 && $\bf{0.84}$     		 && $\bf{0.54}$      && $\bf{-0.90}$      &&  $\bf{-1.52}$       \\
			& $[-4.10, \phantom{0}1.24]$      	 && $[-2.09, \phantom{0}1.77]$     	 && $[-0.47, \phantom{0}4.58]$      	 && $[-1.47, \phantom{0}3.14]$    		
			&& $[-1.65, \phantom{0}2.57]$      && $[-3.18, \phantom{0}1.27]$      && $[-4.06, \phantom{0}0.86]$        \\
			Average in March         & $\bf{\phantom{0}2.68}$      		 && $\bf{1.77}$      				 && $\bf{5.24}$      				 && $\bf{4.51}$     			 && $\bf{2.68}$      && $\bf{2.47}$      && $\bf{\phantom{-}1.30}$        \\
			& $[\phantom{0}0.52, \phantom{0}4.78]$   && $[\phantom{0}-0.41, \phantom{0}3.88]$ && $[\phantom{0}2.33, \phantom{0}7.88]$ && $[\phantom{0}2.01,\phantom{-}7.00]$     
			&& $[\phantom{0}0.19, \phantom{0}5.02]$      && $[\phantom{-}0.36, \phantom{0}5.14]$      && $[-1.00, \phantom{0}3.43]$       \\
			Average in April         & $\bf{9.24}$      					 && $\bf{10.24}$      				 && $\bf{9.47}$      				 && $\bf{10.20}$     			 && $\bf{\phantom{0}9.44}$      && $\bf{7.72}$      && $\bf{\phantom{0}6.36}$    \\
			& $[\phantom{0}6.64, 11.72]$      	 && $[\phantom{0}7.88, 12.66]$ && $[\phantom{0}6.26, 12.32]$      	 && $[\phantom{0}7.26, 12.91]$
			&& $[\phantom{0}6.74, 12.07]$      && $[\phantom{0}4.49, 10.71]$      && $[\phantom{0}3.77, \phantom{0}8.80]$    \\
			Average in May       & $\bf{6.46}$  	 && $\bf{10.71}$      				 && $\bf{10.44}$    	 && $\bf{10.47}$ 		 && $\bf{\phantom{0}7.35}$      && $\bf{9.24}$      && $\bf{\phantom{0}4.44}$    \\
			& $[\phantom{0}3.24, \phantom{0}9.35]$      	 && $[\phantom{}8.28, 13.16]$ && $[\phantom{0}6.70, 13.90]$      	 && $[\phantom{0}7.17, 13.54]$
			&& $[\phantom{0}4.45, 10.20]$      && $[\phantom{0}6.22, 12.07]$      && $[\phantom{0}2.10, \phantom{0}6.59]$ 
			\\
			Average in June      & $\bf{0.29}$  	 && $\bf{3.49}$      				 && $\bf{1.79}$    	 && $\bf{5.72}$ 		 && $\bf{\phantom{0}0.14}$      && $\bf{2.66}$      && $\bf{\phantom{0}2.41}$    \\
			& $[-2.74, \phantom{0}3.04]$      	 && $[\phantom{}1.44, \phantom{0}5.54]$ && $[-1.78,\phantom{0} 5.06]$      	 && $[\phantom{0}2.37,\phantom{0} 8.78]$
			&& $[-2.57,\phantom{0} 2.52]$      && $[-0.05, 5.17]$      && $[\phantom{0}0.54,\phantom{0} 4.06]$ \\
			\bottomrule
		    \toprule
			Electricity Consumption Reduction (\%) & Boston && Chicago && Houston && Kansas City && Los Angeles && New York City && Philadelphia \\
			\midrule
			Average in February      &  $\bf{0.40}$    &&$\bf{0.09}$ && $\bf{-0.55}$ && $\bf{0.10}$	 && $\bf{-1.12}$   &&	$\bf{0.43}$	 && $\bf{0.75}$   \\
			& $[-1.93, \phantom{0}2.60]$ &&$[-2.41, \phantom{0}2.43]$      	 && $[-3.02, \phantom{0}1.93]$     	 && $[-2.76, \phantom{0}2.89]$      	 && $[-4.27,\phantom{-}1.83]$    	&& 	$[-2.12,\phantom{-} 2.90]$
			&& $[-1.98,\phantom{-}3.40]$            \\
			Average in March         &$\bf{\phantom{0}7.12}$   &&$\bf{\phantom{0}2.95}$  && $\bf{-0.53}$    && $\bf{0.24}$ 	 && $\bf{3.32}$&&	$\bf{5.27}$	 && $\bf{3.94}$    \\
			& $[\phantom{0}4.63, \phantom{0}9.53]$ && $[\phantom{0}0.26,\phantom{0}5.49]$   && $[\phantom{-}3.01, \phantom{0}1.70]$ && $[-3.44, \phantom{0}3.57]$ &&   $[\phantom{0}0.61, \phantom{0}5.85]$  && $[2.60, \phantom{0}7.80]$
			&& $[-0.96, \phantom{0}6.86]$   \\
			Average in April         &$\bf{11.32}$  &&$\bf{9.81}$ 	 && $\bf{5.33}$	 && $\bf{9.04}$	 &&  $\bf{11.06}$   	&&	$\bf{14.10}$ 	 && $\bf{8.93}$\\
			&$[\phantom{0}8.55, 13.93]$   &&$[\phantom{0}6.70, 12.66]$  && $[\phantom{0}2.63, \phantom{0}7.79]$ && $[\phantom{0}5.00, 12.55]$  &&  $[\phantom{0}8.11, 13.82]$ && $[11.26, 16.80]$ 
			&& $[\phantom{0}5.42, 12.18]$    \\
			Average in May     &$\bf{9.36}$  &&$\bf{9.51}$ && $\bf{3.63}$ && $\bf{7.01}$   &&  $\bf{3.91}$ 	&&	$\bf{14.77}$ 	 && $\bf{8.24}$\\
			&$[\phantom{0}6.02,12.41]$   &&$[\phantom{0}6.32, 12.51]$      	 && $[\phantom{0}0.86, \phantom{0}5.85]$ && $[\phantom{0}3.22, 10.67]$      	 &&  $[\phantom{0}0.59, \phantom{0}7.06]$ && $[11.61, 17.76]$ 
			&& $[\phantom{0}4.58, 11.71]$    \\
			Average in June         &$\bf{0.41}$  &&$\bf{3.24}$  && $\bf{4.41}$  && $\bf{0.21}$    &&  $\bf{-1.90}$   	&&	$\bf{11.07}$ 	 && $\bf{2.07}$\\
			&$[-3.03,\phantom{0}3.38]$   &&$[\phantom{0}0.36, \phantom{0}5.84]$      	 && $[\phantom{0}2.05, \phantom{0}6.48]$ && $[-2.56, \phantom{0}2.62]$      	 &&  $[-5.42, \phantom{0}1.34]$ && $[\phantom{0}7.60, 14.02]$ 
			&& $[-1.20, \phantom{0}5.06]$    \\
			\bottomrule
		\end{tabular}
		Note: The regional transmission organizations are listed in an order from the Federal Energy Regulatory Commission, and the cities are given in an alphabetical order.
	\end{threeparttable}}
		\linespread{1}
		\captionof{figure}{(a) Electricity consumption profile comparison in NYISO between the ensemble backcast estimations, past profile and real profile. Four typical Mondays are chosen for comparison during February to April. The ensemble backcast estimations include both a point- and interval-estimation, and the $10\%-90\%$, $25\%-75\%$ quantiles are also given. The past electricity consumption profiles in 2019 are aligned with the real profiles by the day of the week. (b) Table showing comparison of changes in electricity consumption across electricity markets and cities in the U.S. All markets and cities experienced a reduction in electricity consumption in April but with diverse magnitude. Particularly, dense urban areas suffered the most severe reduction in April.}
	\label{fig-backcast}
\end{minipage}

\vspace{1em}
\noindent and shutdown of commercial activity slow down the spread of COVID-19. Conversely, a rise in the number of COVID-19 cases results in an increase in social distancing (size of the stay-at-home population), as well as shut down of businesses (commercial loads). This trend is clearly discernible in {mobile device} location data as an increase in {the} stay-at-home population (Supplementary Fig. S-3) and a reduction in visits to retail establishments (Supplementary Fig. S-4). Fig.~\ref{fig-dyn-fac-sd}-c  shows the trace of the evolution of daily new confirmed cases and social distancing, and the associated rate of reduction in electricity consumption for two representative metropolises - New York City and Philadelphia, indicating a fast developing period in March 2020 and a 

\noindent more stable period afterwards. {A slight rebound in the electricity consumption that may be correlated with the partial reopening of the economy, and the relaxation of some social distancing restrictions, is also observed in the trace of the electricity consumption during June 2020.}  Similar trends are observed in other COVID-19 hotspot cities that are in various stages of evolution of the pandemic (Supplementary Fig. S-5). { An alternative visualization of the same result is shown in Supplementary Fig. S-6 for all metropolises.} The trace of evolution of electricity consumption demonstrates the dynamically evolving, multi-dimensional relationship between the number of COVID-19 cases, the size of the stay-at-home population, and the reduction in electricity consumption.

Second, these influencing factors exhibit very different temporal dynamics. For example, in New York City, Fig. \ref{fig-dyn-fac-sd}-b shows a wide variation in the time scales of the changes in the electricity consumption, public health, stay-at-home, work-on-site, and retail mobility data. The mobility in the retail sector has the earliest response  in terms of the rate of change (gradually dropping from late February 2020 and continuing to go down until late April 2020), resulting from bottom-up responses of consumers to the emerging pandemic. On the other hand, the population of on-site workers shows a sharp, abrupt change right around mid-March, as a result of top-down federal and state-level policy decisions such as stay-at-home orders. This insight, to our best knowledge, is first revealed in Fig. \ref{fig-dyn-fac-sd}-b, and suggests a very different efficacy of social distancing arising from top-down government policies and from bottom-up individual responses. Finally, the electricity consumption shows a delayed reduction with respect to the number of COVID cases.

Taking into account these two features, we rigorously quantify the multi-dimensional relationship shown in Fig.~\ref{fig-dyn-fac-sd}-a by calibrating several city-specific Restricted Vector Autoregression (Restricted VAR)~\cite{sims1980macroeconomics} models. Restricted VAR models are powerful tools for multivariate time series analysis with complex correlations, and have been widely adopted in econometrics~\cite{stock2001vector} and electricity markets~\cite{yixian2018vector}. 
Compared with ordinary regression analysis, the Restricted VAR model allows for dependencies between model variables that are too complex to be fully known~\cite{stock2001vector}. Please refer to the Methods section for the definition, and Supplementary Methods SM-1,2,3 for details on the calibration and validation of the Restricted VAR model{, and Supplementary Tables ST-1 and ST-2 for the model parameters, and results of statistical tests on the model}. We now examine the restricted VAR model using the variance decomposition and impulse response analyses as described in Supplementary Method SM-4. The variance decomposition analysis indicates the influencing factors that contribute to changes in electricity consumption, while the impulse response analysis describes the dynamical evolution of the reduction in electricity consumption that would result from a unit shock (1\% increase or decrease) in one influencing factor. We note that the restricted VAR model can be further fine-tuned by selecting the most significant influencing parameters {(see Supplementary Note SN-5 regarding the choice of VAR model parameters, and Supplementary Method SM-5 for the VAR model selection procedure)}. Fig.~\ref{fig-var} and Supplementary Fig. S-7 present the variance decomposition and impulse response analyses for various COVID-19 hotspot cities, indicating the ``delayed'' impact of various influencing factors on electricity consumption. By analyzing Fig.~\ref{fig-var} and Supplementary Fig. S-7, we obtain three key findings.

The first key finding is that the mobility in the retail sector is the most significant and robust factor influencing the decrease in electricity consumption across all cities. This factor accounts for a significant proportion of the change in electricity consumption in both the variance decomposition results (Figs.~\ref{fig-var}-a,c,e) and the impulse response analyses (Figs.~\ref{fig-var}-b,d,f). For example, in Houston, a 1\% decrease in the mobility of retail sector results in a 0.78\% reduction in electricity consumption in the steady state. Further, from the impulse response analyses (Figs.~\ref{fig-var}-b,d,f and Supplementary Figs. S-7-b,d,f,h), the electricity consumption is typically most sensitive to changes in the mobility in the retail sector.

The second finding is that the number of new confirmed COVID-19 cases, although easy to obtain, may not be a strong direct influence on the change in electricity consumption. This finding is supported by observations of a low sensitivity of the electricity consumption to this factor in impulse response results across all cities Figs.~\ref{fig-var}-b,d,f. Note that a high proportion of a particular factor in the variance decomposition may not always mean a high sensitivity to that factor in the impulse response analysis; therefore, the variance decomposition analysis alone cannot be used to infer the magnitude of influence of dependent or correlated influencing factors \cite{lanne2014generalized}. The low sensitivity of the electricity consumption to the number of COVID-19 cases in the impulse response analysis, taken together with its occurrence as an important influencing factor in the variance decomposition, indicates that it exerts an indirect influence on the electricity consumption through other influencing factors (such as social distancing and commercial activity). This result also partly explains the sharp corner in the trace of New York City's electricity consumption in Fig. \ref{fig-dyn-fac-sd}-c after mid April, where no immediate growth in the electricity consumption is observed despite the decrease in number of daily new confirmed cases.

The third finding is that high sensitivities to some influencing factors may be observed in cities with a mild overall reduction in electricity consumption. For example, Fig.~\ref{fig-var}-f indicates that the change in electricity consumption in Houston is very sensitive to variations in the level of commercial activity (mobility in the retail sector), despite the magnitude of the change in electricity consumption not being very significant (Fig.~\ref{fig-backcast}-b). Therefore, such cross-domain insights that are not readily available from traditional analyses may need to be considered in evaluating policy decisions pertaining to the electricity sector. In summary, our findings quantify the dynamics of the interplay between the rise in the number of COVID-19 cases, increased social distancing, and reduced commercial activity, in influencing electricity consumption in the U.S. 
\noindent\begin{minipage}{\textwidth}
    	\includegraphics[width=\textwidth]{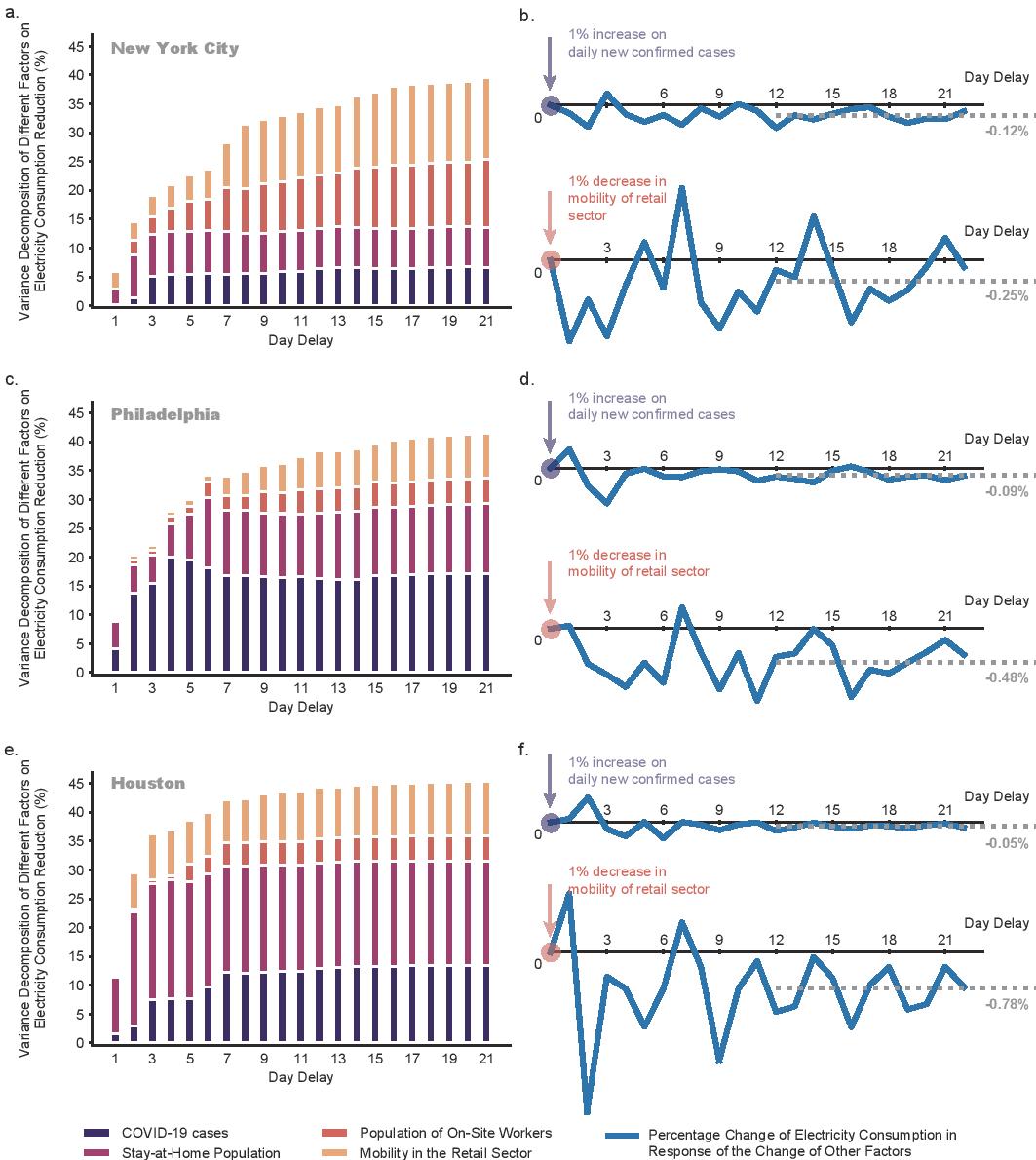}
	\captionof{figure}{Restricted vector autoregression (VAR) model analyses for New York City, Philadelphia and Houston. (a)(c)(e)  Variance decomposition (excluding the inertia of the electricity consumption itself) indicating the contribution of different influencing factors, namely, the daily new confirmed COVID-19 cases, the stay-at-home population and the population of on-site workers (indicative of social distancing),  and mobility in the retail sector (indicative of commercial electricity loads), to changes in electricity consumption. 
	(b)(d)(f) Dynamical evolution of the reduction in electricity consumption that would result from a unit shock (1\% increase or decrease) in one influencing factor.}
	\label{fig-var}
\end{minipage}
\clearpage
\section*{Discussion}
We introduced a timely open-access easy-to-use data-hub aggregating  multiple data sources for tracking and analyzing the impact of COVID-19 on the U.S. electricity sector. The hub will allow researchers to conduct cross-domain analysis on the electricity sector during and after this global pandemic. We further provided the first assessment results with this data resource to quantify the intensity and dynamics of the impact of COVID-19 on the U.S. electricity sector. This research departs from conventional power system analysis by introducing new domains of data that would have a significant impact on the behavior of electricity sector in the future. Our results suggest that the U.S. electricity sector, and particularly the Northeastern region, is undergoing highly volatile changes. The change in the overall electricity consumption is also highly correlated with cross-domain factors such as the number of COVID-19 confirmed cases, the degree of social distancing, and the level of commercial activity observed in each region, suggesting that the traditional landscape of forecasting, reliability and risk assessment in the electricity sector will now need to be augmented with such cross-domain analyses in the near future. We also find very diverse levels of impact in different marketplaces, indicating that location-specific calibration is critically important. 

{ The cross-domain analysis of the electricity sector presented here can immediately inform both power system operators and policy makers as follows. Power system operators can leverage the analysis for short-term planning and and operation of the grid, including load forecasting, and rigorous quantitative assessments of impacts like renewable energy curtailment\cite{caiso1,caiso2} during COVID-19. From a policy-making perspective, the Restricted VAR analysis can be exploited to infer both the key influencing factors such as the mobility in the retail sector that may not be apparent from conventional analyses, and the varied time scales of top-down (policy-level) and bottom-up responses (individual-level), driving changes in electricity consumption. 

This work also opens up several directions for future research by incorporating cross-domain data into the analysis of the electricity sector. For example, vulnerable populations like low-income households are facing an increased energy burden due to COVID-19 \cite{graff2020covid,xu2019energy}. In this context, we are exploring the integration of socio-economic data on demographics \cite{demographic_data} and the social vulnerability index (SVI) \cite{svi_data} into the COVID-EMDA$^+$ data hub. The new cross-domain sources and analysis can then be leveraged by policy-makers to infer the energy burden on such vulnerable populations. The change in electricity consumption can also be an early indicator of the economic impacts of COVID-19 that are not yet reflected in traditional economic indicators like the GDP growth rate. Historically, there has been a significant correlation between electricity use and economic growth over the last four decades \cite{ferguson2000electricity,arora2016electricity,hirsh2015electricity}. Leveraging the cross-domain COVID-EMDA$^+$ data hub, the changes in the electricity sector may be used by policy makers to provide short-term forecasts of the economic impact of COVID-19, including the GDP growth rate, and the level of commercial and industrial activity. The cross-domain Restricted VAR analysis can also be extended to analyze the impact of various policy decisions on the electricity sector, and consequently, the short-term economic health.}
\clearpage

\section*{Experimental Procedures}
\subsection*{Resource Availability}
\subsubsection*{Lead Contact}
Further information and requests for resources and materials should be directed to and will be fulfilled by the Lead Contact, Le Xie (le.xie@tamu.edu)
\subsubsection*{Materials Availability}
No materials were used in this study.

\subsubsection*{Data and Code Availability}
The COVID-EMDA$^+$ data hub and codes for all the analyses in this paper are publicly available on Github~\cite{RN1}. The supporting team will collect, clean, check and update the data daily, and provide necessary technical support for unexpected bugs. In the Github repository, the processed data {(CSV format)} are shared along with the original data {(CSV format)} and their corresponding parsers (written in Python). Several simple quick start examples are included to aid beginners. {The details of the original sources are shown in Supplementary Note SN-4.}

\subsection*{Data Aggregation and Processing Methodology}
In order to obtain cross-domain insights about the impact of COVID-19 on the electricity sector, we integrate data from all U.S. electricity markets with other heterogeneous data like weather, COVID-19 public health, satellite imagery, and {mobile device} location data. The original sources for each dataset are provided in the Data and Code Availability section. Although all seven U.S. electricity markets have established websites for public information disclosure, their download centers, database structures and user interfaces differ widely. Further, file formats, definitions, historical data availability and documentations are also extremely diverse across these markets, making it difficult to integrate this data into a unified framework. The major challenges in integrating data across different electricity marketplaces are as follows.
\begin{itemize}
	\item Some data are stored in hard-to-find pages without {user-friendly} navigation links.
	\item Some data are not packed and collected in an aggregated file for the requested date range. A batch downloader is needed to download these data files one by one, and then aggregate them into the desired single file.
	\item Inconsistent definitions and abbreviations are used among different markets. The same concepts used by different data categories don't follow the same terminology even within the same market.
	\item Geographical information often lacks documentation.
	\item The data quality is not reliable. Data redundancy, duplicate data and missing data are common problems across all markets.
\end{itemize}
As shown in Supplementary Fig. S-1, we design a processing flowchart to reorganize and harmonize all heterogeneous data sources, following three principles - data consistence, data compaction, {and data quality control}, as follows.

\noindent\emph{Data Consistence:} 
\begin{itemize}
	\item For each source of electricity market files, a specific parser is designed to transform the data into a standard long table with date and hour indices. After processing by the parser, raw data from different markets is converted to a unified format.
	\item Geocoding is adopted to match the geographical scale of {electricity market data, COVID-19 case data, and weather data}. 
	\item In the final labeling step, all the field names of data files are translated to the corresponding standard name from a pre-selected name list.
\end{itemize}

\noindent\emph{Data Compaction:} 
\begin{itemize}
	\item Redundant data are dropped by parsers, and the packing step transforms the standard long tables into compact wide tables by pivoting the hour indices as new columns. Usually, the compact wide table can achieve more than 10x file compression rate compared to the un-processed raw files.
	\item COVID-19 cases data are aggregated to the scale of market areas.
	\item The minute-level weather observations are re-sampled into an hourly basis to align with the resolution of market data.
\end{itemize}  

\noindent\emph{{Data Quality Control:}} 
\begin{itemize}
	\item Single missing data (most frequent) are filled by linear interpolation. For consecutive missing data (for example, consecutive missing dates, which are very rare), data from the EIA or EnergyOnline are carefully supplemented.
	\item Outlier data samples are automatically detected when they are beyond $5$ times or below $20\%$ of the associated daily average value. Exceptions such as price spikes and negative prices in LMP data are carefully handled.
	\item Duplicate data are dropped, only the first occurrence of each data sample is kept.
\end{itemize}
{The detailed flow chart of the data quality control used in the COVID-EMDA$^+$ data hub is shown in Supplementary Fig. S-8.}

\subsection*{Ensemble Backcast Model}
The ensemble backcast model is used to estimate the electricity consumption profile in the absence of the COVID-19 pandemic, so that the difference between an ensemble backcast model and the actual metered electricity consumption can be used to quantify the impact of the pandemic. A backcast model is expressed as a function that maps potential factors that may affect electricity consumption level, including weather variables (such as temperature, humidity and wind speed), date of year, and economic prosperity (yearly GDP growth rate) to the estimated electricity consumption. Given a group of backcast models, ensemble forecasting is widely recognized as the best approach to provide rich interval information. A group of backcast models for the daily average electricity consumption can be described by
\begin{align}
	\label{eqn-backcast-load}
	\hat{L}_{md} = \frac{1}{N} \sum_{i=1}^N \hat{f}_i (C_{md}, T_{mdq}, H_{mdq}, S_{mdq}, E ), \ \ \forall m, d,
\end{align}
where $C_{md}$ is the calendar information including month, day, weekday and holiday flag, $\hat{L}_{md}$ is the estimated daily average electricity consumption for month $m$ and day $d$, $\hat{f}_i$ is the $i$th backcast model, $T_{mdq}, H_{mdq}, S_{mdq}$ are temperature, humidity and wind speed within the selected quantiles $q$, and $E$ is the estimated GDP growth rate. We typically include $25\%$, $50\%$ (average value), $75\%$ and $100\%$ (maximum) quantiles, and the final inputs should be decided based on the data after extensive testing. With the backcast estimations, the daily reduction in electricity consumption, $r_{md}$, is calculated as follows,
\begin{align}
	\label{eqn-reduction-rate}
	r_{md} = \Big( 1 - \frac{1}{\hat{L}_{md}} \cdot \frac{1}{T} \sum_{t=1}^T L_{mdt} \Big) \times 100 \%, 
	\ \ \forall m, d,
\end{align}
where $T=24$ is the total number of hours in one day, and $L_{mdt}$ is the electricity consumption metered at time $t$ on month $m$ and day $d$. Equation \eqref{eqn-reduction-rate} compares the ensemble backcast and actual electricity consumption results, and can be readily extended to interval estimations by adjusting the ensemble backcast result. 

\noindent{The detailed procedure adopted here for building the ensemble backcast model is as follows:}
\begin{itemize}
    \item {\textbf{Feature Selection}: 
    We select calendar information (year/month/day, weekday/weekend, holiday flag, etc.), weather data (daily average temperature, humidity, wind speed, etc.), and economic conditions (monthly state-wide GDP) as the input features.}
    
    \item {\textbf{Base Model Selection}: We choose a neural network as the base model, and determine the number of layers and the number of neurons in each layer that minimize the training error, through a random search over the hyperparameter space.  Based on this approach, we find that a four-layer fully-connected neural network with ReLU activation function showed the best performance in terms of accuracy and robustness.}
   %
   %
    \item {\textbf{Model Training}: We then create a large group of model candidates by changing the number of neurons in the base model in a pre-defined range. These model candidates are trained individually by randomly sampling the training data, wherein 85\% of the data-points in 2018 and 2020 are randomly selected as training data, while the remaining 15\% are reserved for verification and evaluation of model performance.}
   %
   %
    \item {\textbf{Model Validation}: The performance of each model is measured by testing over the verification dataset, which contains 15\% of the data-points from 2018 and 2020. We calculate the average prediction error of each month and obtain a $1\times12$ vector for each model, and use the $L_2$ norm of that vector as the error metric. This metric prefers those models that have a reasonable prediction accuracy for every month, instead of those that are very accurate in predicting the load for some months and poor in predicting the load for other months. We train 800 different models and the top 25\% models with the lowest error metric are selected for the final ensemble backcast model.}
    
\end{itemize}

\noindent {In contrast to other algorithms that calibrate weather factors\cite{sullivan2015predicting}, our approach (i) possesses a high degree of flexibility in incorporating more potential influencing factors, (ii) has the ability to capture more complicated correlations, and (iii) gives an accurate estimation of not only expected value but also the probability distribution of the forecasted quantity.}

\subsection*{Restricted Vector Autoregression}
Vector autoregression (VAR) \cite{sims1980macroeconomics} is a stochastic process model that can be used to capture the linear correlation between multiple time-series. We model the dynamics of reduction in electricity consumption using a Restricted VAR model of order $p$ as follows:
\begin{align}
    X_t=C+A_1X_{t-1}+\dots+A_pX_{t-p}+E_t,
\end{align}
where
\begin{align}
    A_i= \begin{bmatrix}
     a_{1,1}^i & a_{1,2}^i & \dots & a_{1,n}^i\\
     a_{2,1}^i & a_{2,2}^i & \dots & a_{2,n}^i\\
     \vdots & \vdots & \ddots  &  \vdots\\
     a_{n,1}^i & a_{n,2}^i & \dots & a_{n,n}^i
    \end{bmatrix},\quad
    X_t=\begin{bmatrix}
     x_t^1 \\
     x_t^2 \\
     \vdots \\
     x_t^n
    \end{bmatrix},\quad
    C=\begin{bmatrix}
     c^1 \\
     c^2 \\
     \vdots \\
     c^n
    \end{bmatrix},\quad
    E_t=\begin{bmatrix}
     e_t^1 \\
     e_t^2 \\
     \vdots \\
     e_t^n
    \end{bmatrix},
\end{align}
in which $A_i$ is the regression matrix, $x_t^1$ represents the target output variable at time $t$, namely the reduction in electricity consumption we wish to model, $x_t^2, ..., x_t^n$ represent the selected $n-1$ parameter variables including confirmed case numbers, stay-at-home population, median home dwell time rate, population of on-site workers, mobility in the retail sector and etc., $C$ and $E_t$ are respectively column vectors of intercept and random errors, and the time notation $t-p$ represents the $p$-th lag of the variables. 

\noindent {The full procedure of building the restricted VAR model mainly contains four steps as follows, including pre-estimation preparation, restricted VAR model estimation, restricted VAR model verification, and post-estimation analysis. These steps, outlined below, are detailed in Supplementary Methods SM-1 to SM-5.}
\vspace{0.5em}

\noindent\emph{{Pre-estimation Preparation}:}

\begin{itemize}
    \item {Data Preprocessing: Several datasets are collected to calculate the inputs of restricted VAR model, including electricity market data, weather data, number of COVID-19 cases, and mobile device location data. We take logarithms of several variables, including electricity consumption reduction, new daily confirmed cases, stay-at-home population, population of full-time on-site workers, population of part-time on-site workers, and mobility in the retail sector, while only keeping the original value of the median home dwell time rate.}
    \item {Augmented Dickey-Fuller (ADF) Test: Test whether a time-series variable is non-stationary and possesses a unit root.}
    \item {Cointegration Test: Test the long-term correlation between multiple non-stationary time-series.}
    \item {Granger Causality Wald Test: Estimate the causality relationship among two variables represented as time-series.}
\end{itemize}

\noindent\emph{{Restricted VAR Model Estimation}:}

\begin{itemize}
    \item {Ordinary Least Square (OLS): We impose constraints on the OLS to eliminate any undesirable causal relationships between variables.}
\end{itemize}

\noindent\emph{{Restricted VAR Model Verification}:}

\begin{itemize}
    \item {ADF Test: Verify if the residual time-series are non-stationary and possess a unit root.}
    \item {Ljung-Box Test: Verify the endogeneity of the residual data that may render the regression result untrustworthy.}
    \item {Durbin-Watson Test: Detect the presence of autocorrelation at log $1$ in the residuals of the Restricted VAR model}
    \item {Robustness Test: Test the robustness of the Restricted VAR model against parameter perturbations.}
\end{itemize}

\noindent\emph{{Post-estimation Analysis}:}

\begin{itemize}
    \item {Impulse Response Analysis: Describe the evolution of the Restricted VAR model's variable in response to a shock in one or more variables.}
    \item {Forecast Error Variance Decomposition: Aid in the interpretation of the Restricted VAR model by determining the proportion of each variable's forecast variance that is contributed by shocks to the other variables.}
\end{itemize}


\Urlmuskip=0mu plus 1mu\relax  
\clearpage

\flushbottom
\maketitle
\thispagestyle{empty}
\setcounter{figure}{0}  
\section*{Introduction}
This document presents the detailed analysis procedures and additional results to supplement the main text, and is organized into three sections as follows: 1) supplementary figures containing analysis results (using the same analysis and visualization methods) for several cities not included in the main body; 2) supplementary notes introducing the definitions and processing of cross-domain datasets; 3) supplementary methods presenting the statistical analysis used in establishing and evaluating the restricted VAR models. The contents of this document are listed as follows:
\subsection*{Supplementary Figures:}
\begin{itemize}
	\item Fig. S-1: Architecture of COVID-EMDA$^+$ data hub
	\item Fig. S-2: Night Time Light Images in COVID-19 Hotspot Cities
	\item Fig. S-3: Comparison of Stay-at-home Population
	\item Fig. S-4: Change of Visits to Points of Interest
	\item Fig. S-5: Visualization of COVID-19 Cases, Stay-at-home Population, and Reduction in Electricity Consumption
	\item {Fig. S-6: Alternative Visualization of COVID-19 Cases, Size of Stay-at-home Population, and Reduction in Electricity Consumption}
	\item Fig. S-7: Additional VAR Results 
	\item {Fig. S-8: Details of Data Quality Control}
	\item {Fig. S-9: Google Trends Activity Data of COVID-19 Related Keywords}
	\item {Fig. S-10: Trend Transition of Cross-domain Variables}
	
\end{itemize}
\subsection*{Supplementary Notes:}
\begin{itemize}
	\item SN-1: Description of the Night-Time Light Dataset
	\item SN-2: Description of the Mobile Device Location Dataset
	\item SN-3: Definition of "Retail" Data Used in the restricted VAR Model
	\item {SN-4: Data Sources for the COVID-EMDA$^+$ Data Hub}
	\item {SN-5: Remarks on Choice of Restricted VAR Model Parameters - Cases vs. Hospitalizations/Deaths}
\end{itemize}

\subsection*{Supplementary Methods:}
\begin{itemize}
	\item SM-1: Pre-estimation Preparation
	\item SM-2: Restricted VAR Model Estimation
	\item SM-3: Restricted VAR Model Verification
	\item SM-4: Post-estimation Analysis
	\item SM-5: Restricted VAR Model Selection
\end{itemize}
\subsection*{{Supplementary Table:}}
\begin{itemize}
	\item {ST-1: Restricted VAR Model Parameters}
	\item {ST-2: Restricted VAR Statistical Test Results}
	\item {ST-3: Availability and Correlation Test of COVID-19 Public Health Metrics}
	\item {ST-4: Statistical Tests of Restricted VAR Models Using Different COVID-19 Indicator Variables}
\end{itemize}

\clearpage
\section*{Supplementary Figures}
\subsection*{Supplementary Figure S-1: Architecture of COVID-EMDA$^+$ data hub}
\begin{figure}[!h]
	\centering
	\includegraphics[width=0.98\textwidth]{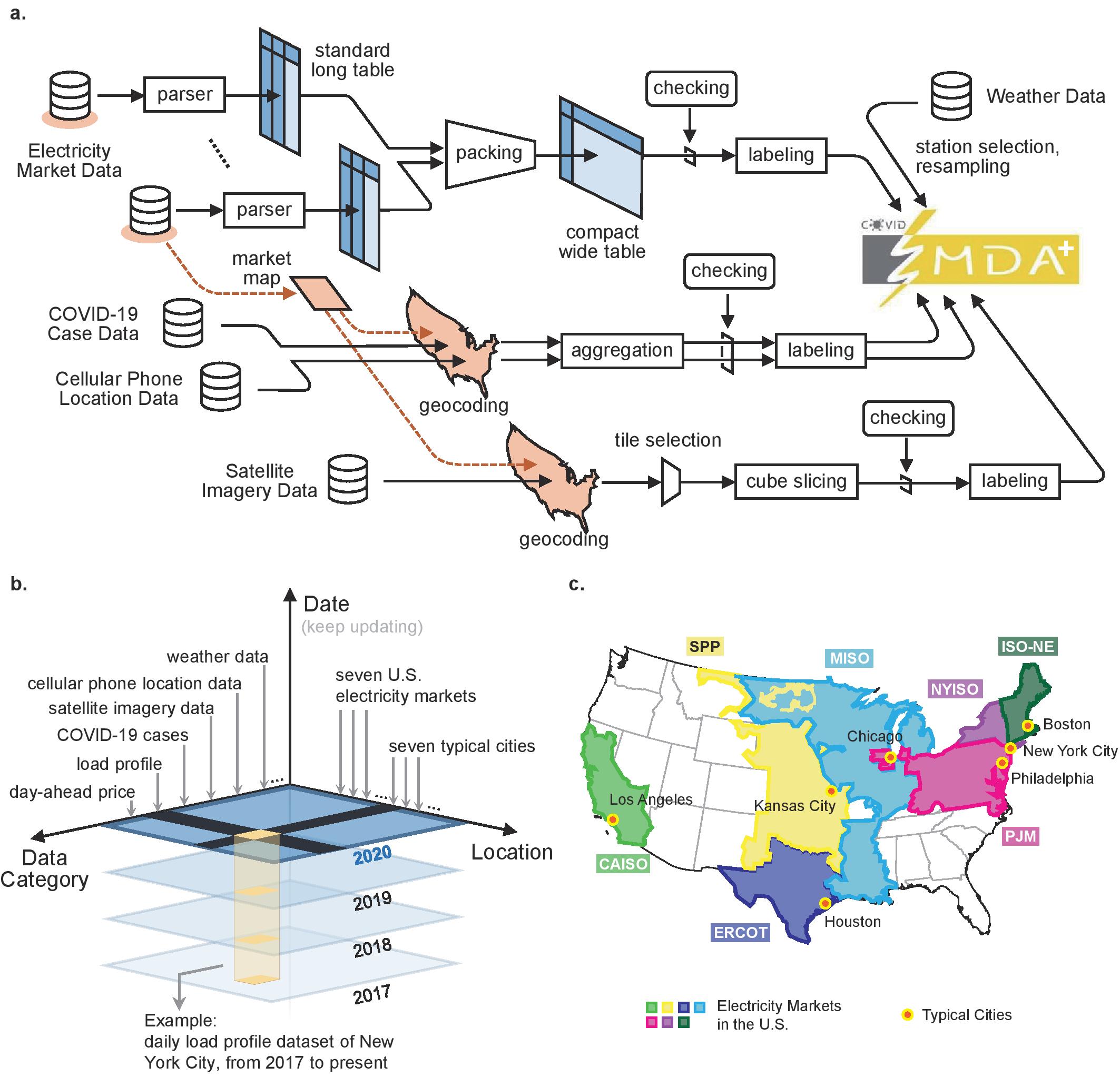}
	\caption{Architecture of COVID-EMDA$^+$ data hub. (a) Processing flowchart of COVID-EMDA$^+$ data hub. Heterogeneous data sources are handled, including electricity market, COVID-19 cases, mobile device location data, satellite imagery and weather data. To coordinate five data sources in the same geographical scales, the geocoding technique is applied to transform COVID-19 cases and weather data. The entire processing reflects the objective of data consistence, data compaction and data checking. (b) The architecture contains the date, data category and location dimensions. The main dimension is the date due to the importance of time-series relationships. Along the main dimension, one can retrieve multiple data slices or spreedsheet data files. The yellow cubic represents one such load dataset for New York City. (c) Map of the United States representing the regions of operation of the seven RTOs or electricity markets.
	}
	\label{fig-datahub-arch}
\end{figure}

Fig.~\ref{fig-datahub-arch} shows the architecture of COVID-EMDA$^+$ data hub, which cross-references information across three categories, namely, different dates, data types (electricity market, weather, public health, mobile device location, and satellite imagery), and locations (RTOs or representative cities). Details on data pre-processing and data quality monitoring are described in the Methods section.

\subsection*{Supplementary Figure S-2: Night Time Light Images in COVID-19 Hotspot Cities}
Fig. S-\ref{fig-ntl} shows the reduction in night-time light brightness, providing a visual representation of the effect of COVID-19 on electricity consumption level in major cities, as the drop in light intensity is obvious and significant.
\begin{figure}[htbp!]
	\centering
	\includegraphics[width=0.98\textwidth]{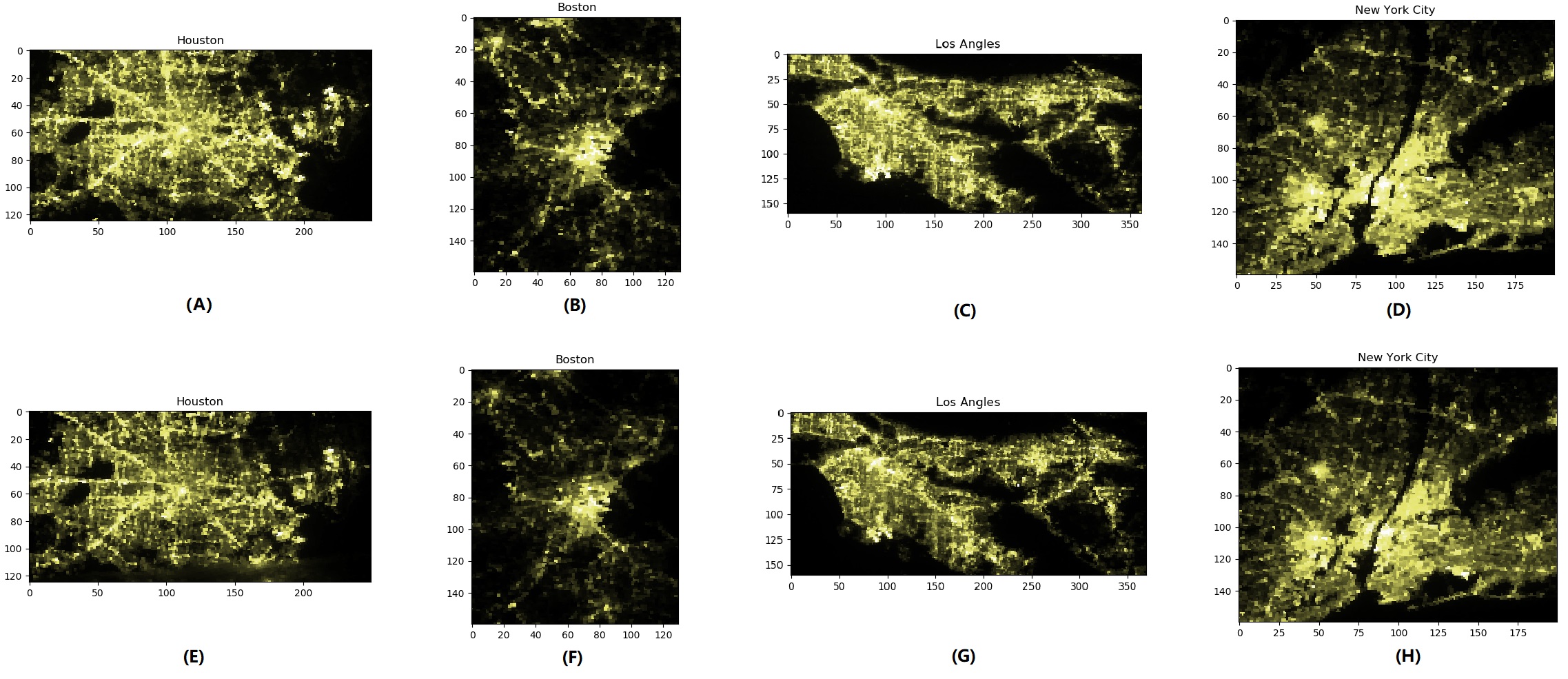}
	\caption{NTL data of 4 major metropolises in the United States. Sub-figures (A) - (D) show the night-time light images before the outbreak of COVID-19 (Early and mid February); (E) - (H) show the nighttime light images during the pandemic (Late April).}
	\label{fig-ntl}
\end{figure}
\clearpage

\subsection*{Supplementary Figure S-3: Comparison of Stay-at-home Population}
\begin{figure}[htbp!!]
	\centering
	\includegraphics[width=0.98\textwidth]{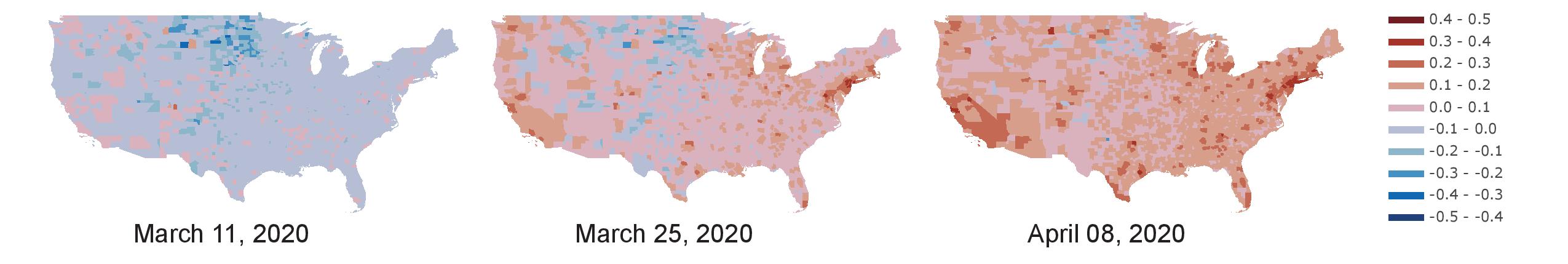}
	\caption{Increased proportion of stay-at-home population with February $12$ being the baseline. All the selected dates are Wednesdays and non-holidays.}
	\label{fig-social-distancing}
\end{figure}

Fig. S-\ref{fig-social-distancing} depicts a significant increase in the social distancing level indicating the change of people's mobility amidst the pandemic, with regional differences based on stringency and effectiveness of stay-at-home policies. 

\clearpage

\subsection*{Supplementary Figure S-4: Change of Visits to Points of Interest (POIs)}
\begin{figure}[htbp!]
	\centering
	\includegraphics[width=0.95\textwidth]{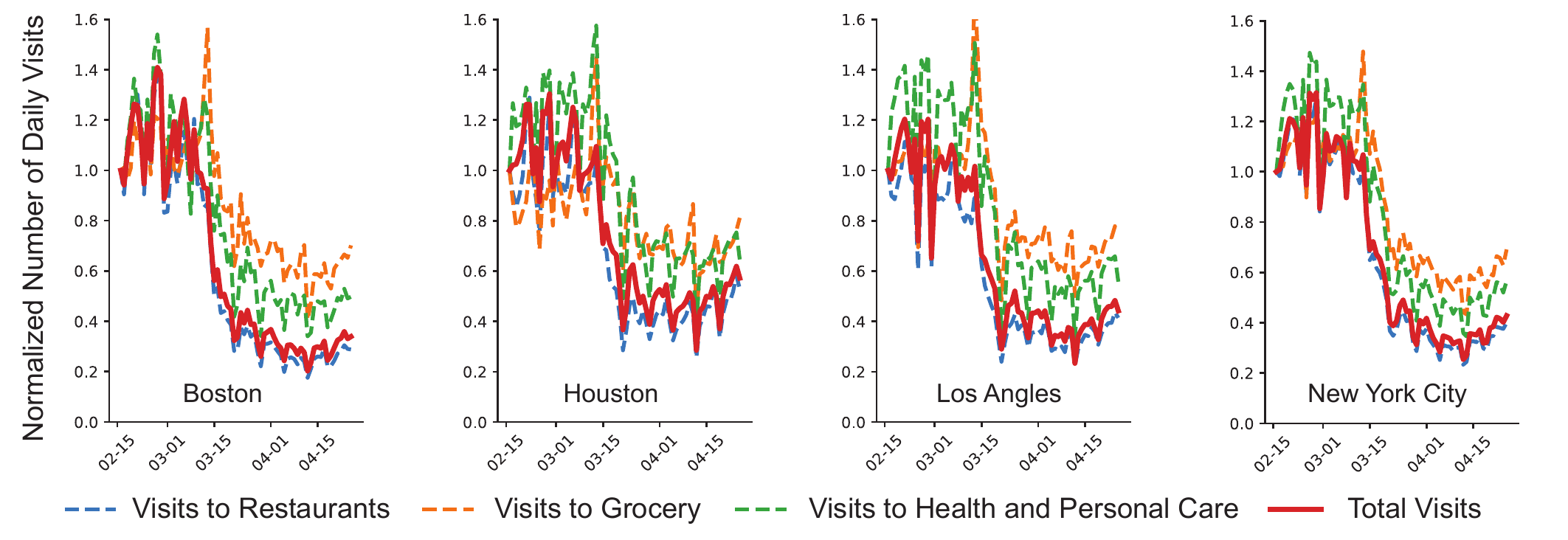}
	\caption{Normalized number of daily total visits and visits to three selected POIs (restaurant, grocery, health and personal care) from February 15 to April $25$, $2020$. The normalized numbers show the relative values of the daily visits with February $15$ being the baseline.}
	\label{fig-pattern}
\end{figure}

Fig. S-\ref{fig-pattern} visualizes the change of visit patterns to common POIs in four hotspot cities from February $15$ to April $25$, from which we can observe that (i) all cities suffered a sudden decline starting from March $13$, the issuing date of the national emergency, and (ii) the extent of the declines have similar characteristics with some regional divergences.

\clearpage

\subsection*{Supplementary Figure S-5: Visualization of COVID-19 Cases, Size of Stay-at-home Population and Reduction in Electricity Consumption}
Fig. S-\ref{fig-covid-power-home} shows the trace of the reduction in electricity consumption, new confirmed COVID-19 cases, and stay-at-home population in Boston, Houston and Kansas City, to supplement the result in Fig. 3-c of the main body.

\begin{figure}[htbp!]
	\centering
	\includegraphics[width=0.95\textwidth]{fig/bubble-type1-5fig.jpg}
	\caption{Trace of the reduction in electricity consumption, new confirmed COVID-19 cases and stay-at-home population in Boston, { Chicago, Los Angeles}, Houston and Kansas City. The bubble sizes indicate the percentage reduction in electrical consumption (with larger bubble sizes indicating more reduction in consumption).
		The number of COVID-19 cases and the size of the stay-at-home population are smoothed by a weekly moving average to properly extract the trends.}
	\label{fig-covid-power-home}
\end{figure}
\clearpage

\subsection*{{Supplementary Figure S-6: Alternative Visualization of COVID-19 Cases, Size of Stay-at-home Population, and Reduction in Electricity Consumption}}
\begin{figure}[htbp!]
	\centering
	\includegraphics[width=0.7\textwidth]{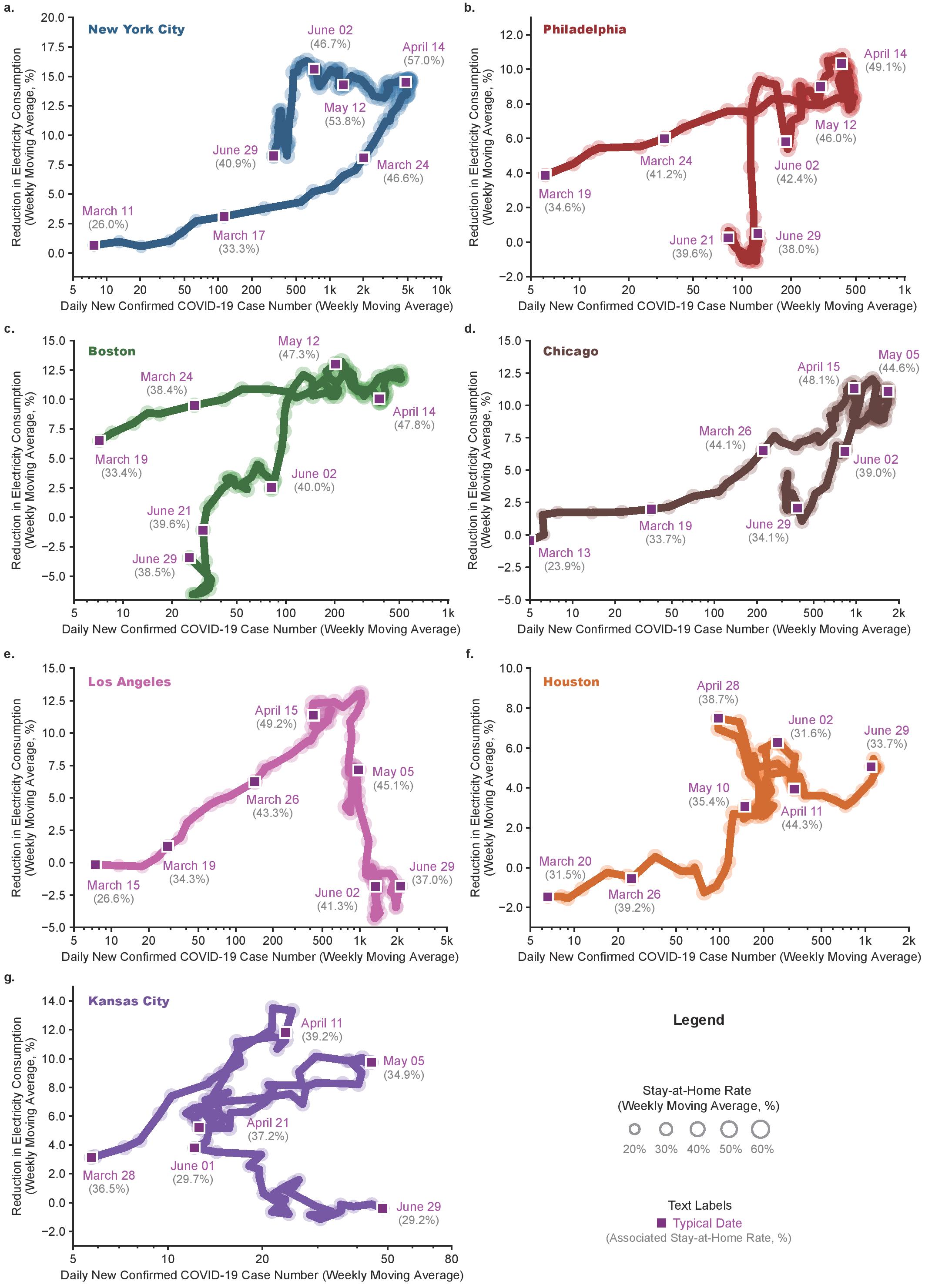}
	\caption{{Trace of the reduction in electricity consumption, number of new confirmed COVID-19 cases, and stay-at-home population in New York City, Philadelphia, Boston, Chicago, Los Angeles, Houston, and Kansas City. The bubble sizes indicate the size of the stay-at-home population (with larger bubble sizes indicating a larger stay-at-home population).
			The number of COVID-19 cases and the size of the stay-at-home population are smoothed by a weekly moving average to properly extract the trends.}}
	\label{fig-covid-power-home-1}
\end{figure}

{Fig. S-\ref{fig-covid-power-home-1} shows an alternative visualization of the trace of the reduction in electricity consumption, new confirmed COVID-19 cases, and stay-at-home population in New York City, Philadelphia, Boston, Chicago, Los Angeles, Houston, and Kansas City, to supplement the result in Fig. 3-c of the main body.}

\clearpage

\subsection*{Supplementary Figure S-7: Additional VAR Analysis Results}
\begin{figure}[htbp!]
	\centering
	\includegraphics[width=0.95\textwidth]{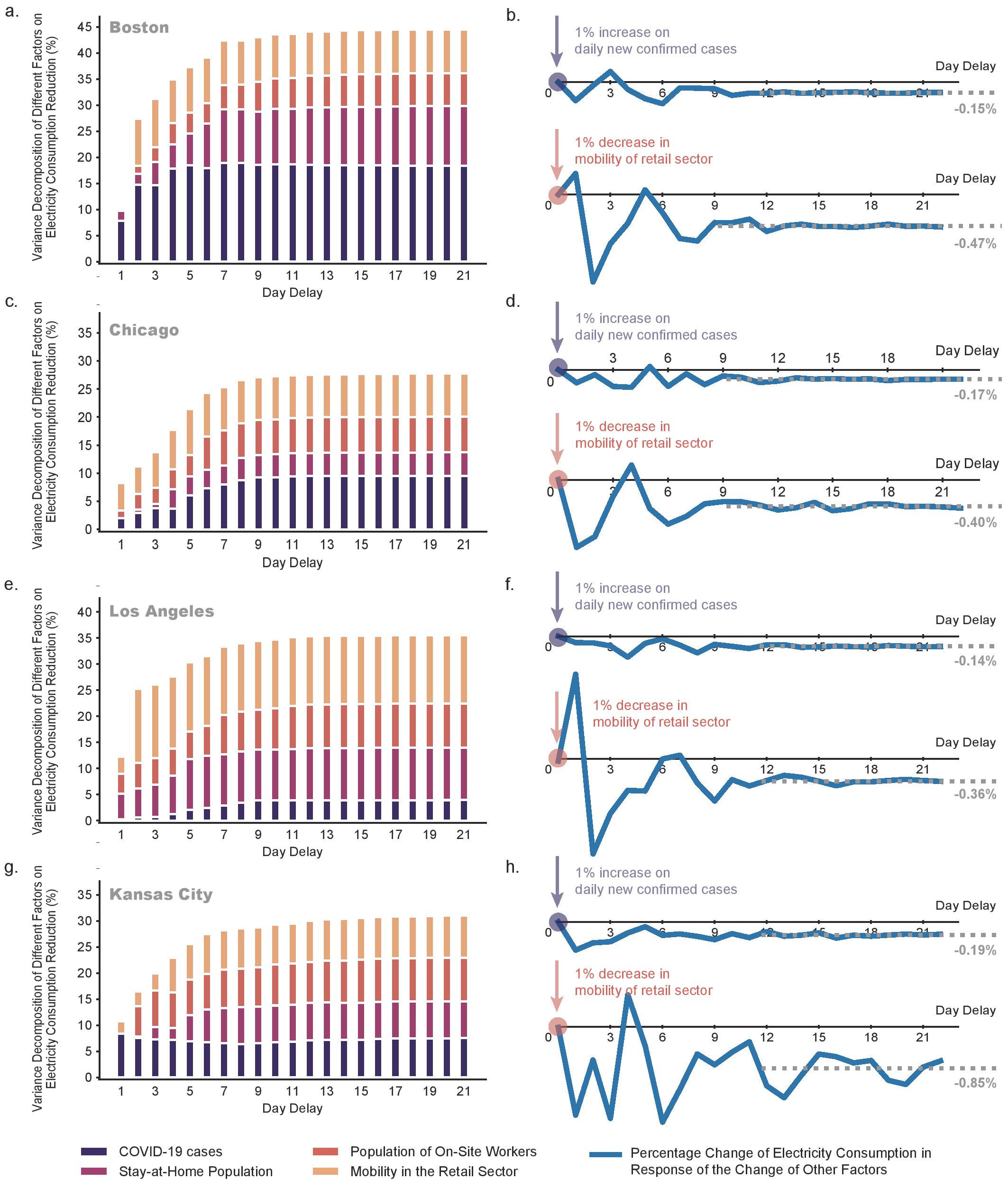}
	\caption{Restricted VAR model analysis for {Boston}, Chicago, Los Angeles and Kansas City. (a)(c)(e){(g)}  Variance decomposition (excluding the inertia of the electricity consumption itself) indicating the contribution of different influencing factors, namely, the daily new confirmed COVID-19 cases, the stay-at-home population and the population of on-site workers (indicative of social distancing),  and mobility in the retail sector (indicative of commercial electricity loads), to changes in electricity consumption. 
		(b)(d)(f){(h)} Dynamical evolution of the reduction in electricity consumption that would result from a unit shock (1\% increase or decrease) in one influencing factor.}
	\label{fig-var_results}
\end{figure}

\clearpage

\subsection*{{Supplementary Figure S-8: Details of Data Quality Control}}
\begin{figure}[htbp!]
	\centering
	\includegraphics[width=0.6\textwidth]{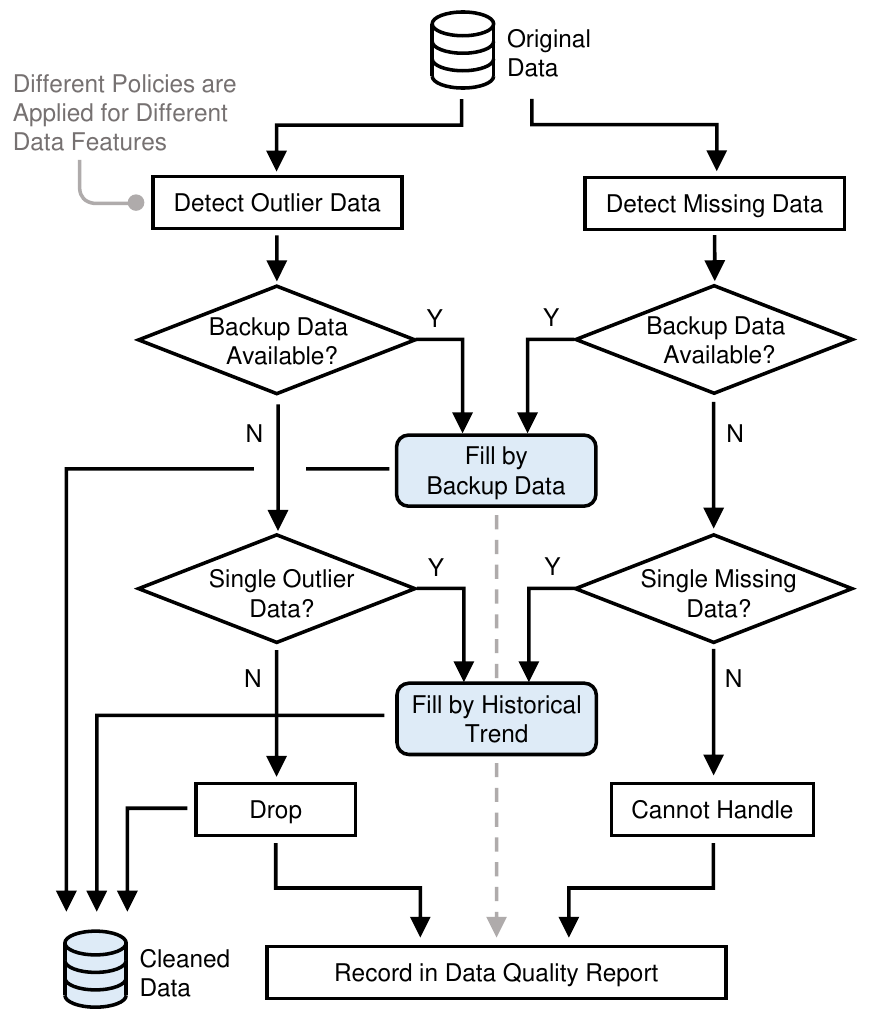}
	\caption{{Data quality control mainly includes two aspects: outlier detection and missing data recovery. Backup data and historical trend are used to achieve both functions. For those that cannot be handled, we record them in the data quality report as shown in the Github as well.}}
	\label{fig-data_quality}
\end{figure}

\clearpage

\subsection*{{Supplementary Figure S-9: Google Trends Activity Data}}
\begin{figure}[htbp!]
	\centering
	\includegraphics[width=0.95\textwidth]{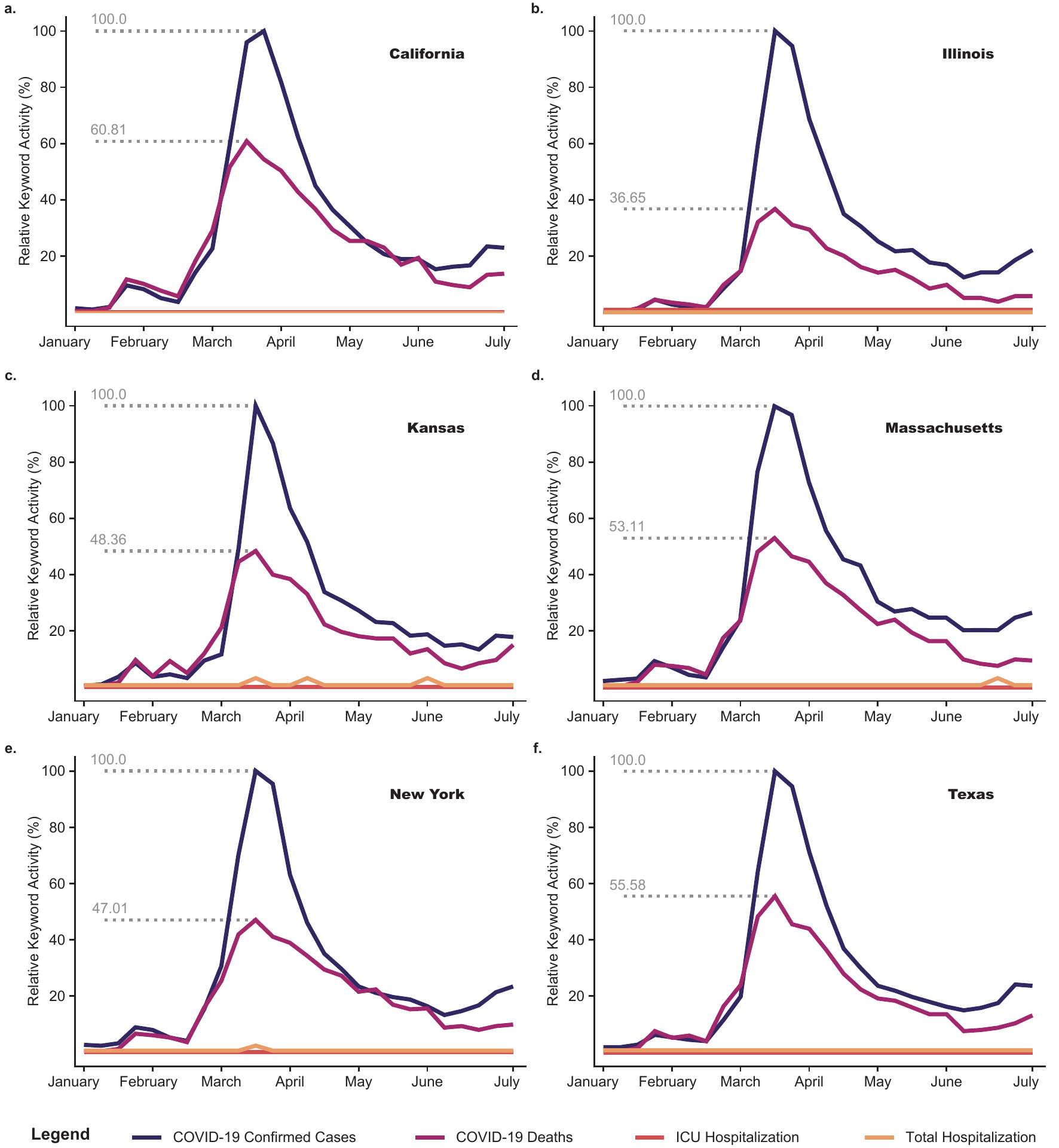}
	\caption{{Relative activity of each category of keywords (indicating the number of COVID-19 cases, number of COVID-19 deaths, number of ICU hospitalizations, and total number of hospitalizations) from Google Trends for each state where the hotspot cities are located, except for Pennsylvania, as we could not access Google Trends data for that state. Each curve shows the total weekly activity of all keywords from one category between 01/01/2020 and 06/28/2020. The list of all keywords associated with each category is collected using the Google Trend feature "relate-queries". The resolution of the Google Trends data is only up to 1\%. The Google search activity associated with both the ICU and hospitalization categories of keywords is typically less than 1\% compared to the activity of the number of new cases, which resulted in both these curves being mostly zero. The plot shows that number of COVID-19 cases and deaths are the two most searched factors, while ICU and hospitalization statistics received very little attention in comparison.}}
	\label{fig-google_trend}
\end{figure}

\clearpage

\subsection*{{Supplementary Figure S-10: Trend Transition of Cross-domain Variables}}
\begin{figure}[htbp!]
	\centering
	\includegraphics[width=0.95\textwidth]{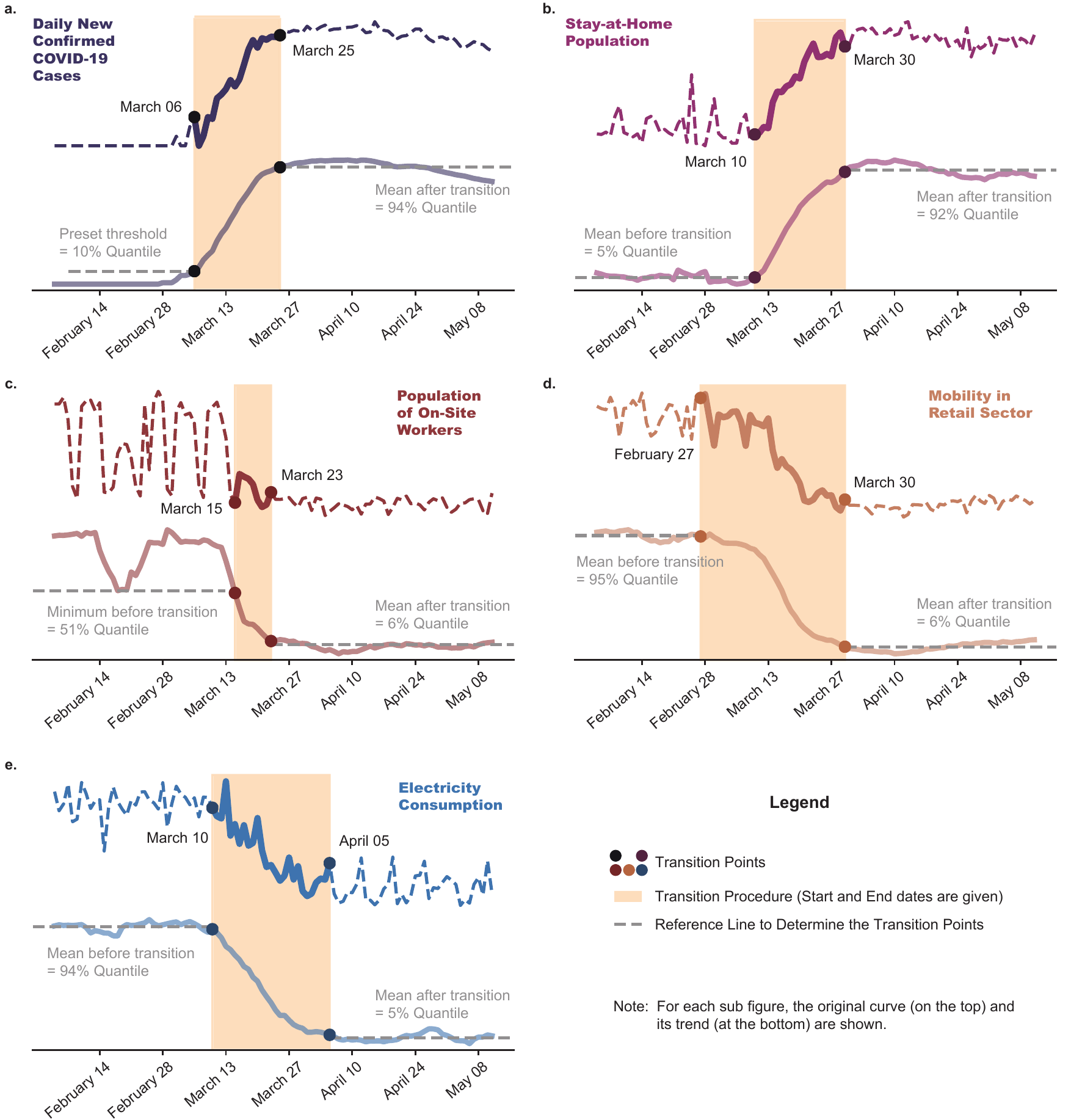}
	\caption{{Trend transition of cross-domain variables. First, we calculate the trend of each variable by eliminating the periodical pattern. The algorithm we choose is tsa.seasonal.seasonal$\_$decompose from the Statsmodels package in Python. The seasonal component is first removed by applying a convolution filter to the data. The average of this smoothed series for each period is the returned trend. Further we make an assumption that there is only one transition period for each variable transferring from one steady stage to another. Based on this assumption, we determine the beginning and end of the transition period. The beginning of the transition period is defined as the latest day on which the trend value is closest to the average value of all the days before. Similarly, we select as the end of the transition period the earliest day on which the trend value is closest to the average value on all following days. Note that the population of on-site workers is one exception in terms of determining the beginning of its transition period, due to one roller-coaster period in the early stages. Therefore, for the population of on-site workers, we determine the beginning of the transition period as the earliest day when the trend value is lower than the previous valley.}}
	\label{fig-trend-transition}
\end{figure}

\clearpage
\section*{Supplementary Notes}
\subsection*{Supplementary Note SN-1: Description of the Night-Time Light Dataset}
Recent progress in on-board sensors and data processing algorithms for remote sensing satellites has opened up many opportunities for monitoring and analyzing human activities on the surface of the Earth and characterizing the impact of human activities on the environment, using satellite data on emission, radiation, atmosphere, vegetation, and water bodies. Among the wide range of available data, Night-Time Light (NTL) has been well recognized as a valuable and unique source of data for understanding the changes in human footprints and economic dynamics\cite{NTLReview}. For our study, the NASA VNP46A1 "Black-Marble" \cite{BlackMarble} dataset is selected as the data source for its high resolution, public availability and daily update. VNP46A1 is collected by the NASA Suomi NPP sun-synchronous remote sensing satellite\cite{Suomi} which has a orbiting period of 101.44 minutes. This satellite measures the surface light radiation at a constant resolution of 500 meter per sample and samples daily at around local time mid-night for every location across the globe. This dataset has been used in power system studies from the perspectives of outage detection\cite{Wang2018Outage} and grid restoration\cite{Roman2019Restoration}.

The NTL dataset is used in this study as a tool for visualizing the impact of COVID-19 on electricity consumption. { We note that we only use the satellite image data for illustration and visualization, but not for numerical analysis, because the sampling frequency of satellite images is too low in comparison to the other data sources we use. Each location is sampled only once per day and most samples are contaminated by the presence of clouds that block the light over the area we are interested in. Further, since a valid and informative satellite image sample must be taken when the sky is mostly clear of cloud, the frequency of valid data is even lower.}

We conduct a comparative study of the impact of COVID-19 on artificial nightlights for representative metropolises in different RTO regions. Specifically, we focus on the cities of Boston, New-York City, Los Angles, and Houston. For each city we select a typical day in both February (before the COVID-19 outbreak) and in April (during the outbreak). The two representative snapshots selected for each city are taken from the same day-of-week and time-of-day, when the sky is clear of cloud.

The raw at-sensor-Day-Night Band (DNB) data is pre-processed using the following procedures to reduce disturbances:
\begin{enumerate}
	\item Manually locate the rectangle containing the targeted city on the tile-level NTL dataset.
	\item Scan the raw data for abnormal pixels (indicated by a pixel-level Quality Flag), and approximate it by taking the average of neighboring pixels.
	\item Scale every pixel using the corresponding moon illumination fraction and pixel-level lunar angles of that day to reduce disturbances from the moon.
	\item Set pixels that have extremely low light intensity (< 10 $nW\cdot cm^{-2}\cdot sr^{-1}$) to 0, to eliminate random ambient noises.
	\item Apply a 5x5 low-pass kernel filter to smooth the image.
	\item Map the light intensity values of each pixel to color using a colormap and plot on axis.
\end{enumerate}
The processed NTL images are presented in Supplementary Fig. S-1. The reduction in night-time light brightness provides a visual representation of the effect of COVID-19 on electricity consumption level in major cities, as the drop in light intensity is obvious and significant.
\clearpage

\subsection*{Supplementary Note SN-2: Description of the Mobile Device Location Dataset}
The original mobile device location dataset is obtained from SafeGraph~\cite{sg_social,sg_pattern}, a data company that aggregates anonymized GPS location data from numerous applications by census block group in order to provide location information. The original dataset contains two major sub-datasets: (i) social distancing metrics and (ii) pattern of visits to Points of Interest (POIs). 
\subsubsection*{Social Distancing Metric}
The social distancing metric dataset is generated using a panel of GPS signals from anonymous mobile devices. Note that "home" is defined as the common nighttime location of each mobile phone over a $6$ week period, and part-time and full-time workplaces are defined as the non-home locations where users  spend from $3-6$ hours and $\geq6$ hours respectively between $8$ am and $6$ pm in local time.

The following are the features selected for our analysis:
\begin{itemize}
	\item \textbf{Basic Information}: (i) unique 12-digit FIPS code for the Census Block Group; (ii) start and end time for the measurement period (namely $24$ hours); (iii) count of devices whose homes are in the Census Block Group.
	
	\item \textbf{Completely Stay-at-home Device Count}: the number of devices that never leave "home" during the measurement period, out of the total count of devices in the Census Block Group.
	
	\item \textbf{Median Home Dwell Time}: the median dwell time at "home" in minutes during the measurement period, for all devices in the Census Block Group.
	
	\item \textbf{Part-time Work-on-Site Device Count}: the number of devices that go to part-time workplaces during the measurement period, out of the total count of devices in the Census Block Group.
	
	\item \textbf{Full-time Work-on-Site Device Count}: the number of devices that go to full-time workplaces during the measurement period, out of the total count of devices in the Census Block Group.
\end{itemize}

We aggregate the daily social distancing data by county. Denote the total count of the completely stay-at-home devices in a county as $C_1$, the median value of the median home dwell time in a county as $C_2$, the total count of the part-time work-on-site devices in a county as $C_3$, the total count of the full-time work-on-site devices in a county as $C_4$ and the total count of devices in a county as $C$. Then, we define $C_1/C$, $C_2/1440$, $C_3/C$, and $C_4/C$ as the county-level "completely stay-at-home rate", "home dwell time rate", "part-time work-on-site rate" and "full-time work-on-site rate" respectively, which will be used for the restricted VAR model.

\subsubsection*{Visit Pattern of POIs}
The pattern dataset is a place traffic and demographic data aggregation available for about $4$ million POIs that contains the frequency of visits to various POIs, the dwell time, the residence location of visitors, etc.

The following are the features selected for our analysis:
\begin{itemize}
	\item \textbf{Basic Information}: (i) the unique and consistent ID tied to each POI; (ii) name of the POI; (iii) physical address; (iv) postal code; (v) brand; (vi) start and end time for measurement period (about one week) in local time; (vii) associated category of the POI.
	
	\item \textbf{Visits by Day}: the number of visits to the POI each day over the covered time period.
\end{itemize}

The daily pattern data for "retail" POIs are collected as defined in Supplementary Note SN-3. We aggregate and sum up the total number of daily visits by county, which will be used as "retail mobility" data in the restricted VAR model.

{ We also  make the following note regarding biases in the SafeGraph data set. SafeGraph officially answers questions about the bias of the dataset \cite{safegraph_bias1,safegraph_bias2} as follows. SafeGraph Patterns measures foot-traffic patterns to 3.6 million commercial points-of-interest from over 45 million mobile devices in the United States, which are about 10$\%$ of devices in the U.S. SafeGraph explores sampling bias on several dimensions including geographic (state/county/census block group) and demographic (race/educational attainment/household income) perspectives. The quantitative analysis results \cite{safegraph_bias2} show that the dataset is nearly unbiased in the mentioned aspects.}
\clearpage

\subsection*{Supplementary Note SN-3: Definition of "Retail" Data Used in Restricted VAR Model}
The "retail" data refers to aggregation of several categories of POIs from the ``pattern of visits to POIs'' dataset\cite{sg_pattern}. We select the following $25$ categories among a total of $168$ POIs as indicators of mobility in the retail sector:
\begin{enumerate}
	\item Automobile Dealers
	\item Automotive Parts, Accessories, and Tire Stores
	\item Beer, Wine, and Liquor Stores
	\item Book Stores and News Dealers
	\item Clothing Stores
	\item Department Stores
	\item Drinking Places (Alcoholic Beverages)
	\item Electronics and Appliance Stores
	\item Florists
	\item Furniture Stores
	\item Gasoline Stations
	\item General Merchandise Stores, including Warehouse Clubs and Super-centers
	\item Grocery Stores
	\item Health and Personal Care Stores
	\item Home Furnishings Stores
	\item Jewelry, Luggage, and Leather Goods Stores
	\item Lawn and Garden Equipment and Supplies Stores
	\item Office Supplies, Stationery, and Gift Stores
	\item Other Miscellaneous Store Retailers
	\item Other Motor Vehicle Dealers
	\item Restaurants and Other Eating Places
	\item Shoe Stores
	\item Specialty Food Stores
	\item Sporting Goods, Hobby, and Musical Instrument Stores
	\item Used Merchandise Stores
\end{enumerate}
\clearpage

\subsection*{{Supplementary Note SN-4: Data Sources for the COVID-EMDA$^+$ Data Hub}}
The original data sources for the COVID-EMDA$^+$ data hub are as follows.
\begin{itemize}
	\item \textbf{Electricity market data:} Data pertaining to load, generation mix and day-ahead locational marginal price (LMP) are obtained from the California (CAISO)~\cite{RN2}, Midcontinent (MISO)~\cite{RN3}, New England (ISO-NE)~\cite{RN4}, New York (NYISO)~\cite{RN5}, Pennsylvania-New Jersey-Maryland Interconnection (PJM)~\cite{RN6}, Southwest Power Pool (SPP)~\cite{RN7}, and the Electricity Reliability Council of Texas (ERCOT)~\cite{RN8}. Electricity market data from the Energy Information Administration (EIA)~\cite{RN9} and EnergyOnline company~\cite{RN10} is used to improve quality and fill in missing data. \\ { \textit{Purpose}: This data has conventionally been used in power system load forecasting and analysis to understand changes in electricity consumption.} 
	\item \textbf{Weather data:}  The original weather data are obtained from the Iowa State University~\cite{RN12}.  Various kinds of weather data (temperature, relative humidity, wind speed and dew temperature) are collected from Automated Surface Observing Systems (ASOS) stations that are supported by the National Oceanic and Atmosphere Administration (NOAA), and can be extracted through an interactive website\cite{RN12}. {Note that typical ASOS stations are selected with consideration of geographical distribution and missing data rate. We select two to three stations for each market area, whose average missing data rates are below $0.5\%$.}
	\\ { \textit{Purpose}: We integrate weather and economic data (state-level seasonally adjusted GDP growth rate) to estimate an accurate baseline electricity consumption profile taking into account weather, calendar, and economic factors, against which the impact of COVID-19 will be quantified.} 
	\item \textbf{Satellite imagery data:} In our study, the NASA VNP46A1 "Black-Marble" \cite{BlackMarble} dataset is selected as the source of satellite imagery data for its high resolution, public availability and daily update. VNP46A1 is collected by the NASA Suomi NPP sun-synchronous remote sensing satelite\cite{Suomi} which has an orbiting period of 101.44 minutes. This satellite measures the surface light radiation at a constant resolution of 500 meter per sample and samples daily at around local time mid-night for every location across the globe.  
	\\ { \textit{Purpose}: We integrate satellite data (NTL data) as a tool for intuitive visualization of the impact of COVID-19 on electricity consumption, especially in large urban and commercial centers.} 
	\item \textbf{COVID-19 public health data:} The original sources of confirmed COVID-19 case numbers and deaths is the John Hopkins University dataset~\cite{RN17}, which contains county-level confirmed case and death numbers from January 22, 2020 (first U.S. case) onwards.
	\\ { \textit{Purpose}: In order to directly assess the impact of the pandemic on electricity consumption, we integrate public health data on COVID-19 cases. However, the increase in COVID-19 cases indirectly influences electricity consumption through factors like social distancing and shut down of commercial activity, that may not be captured solely using COVID-19 case data for electricity consumption analysis.}
	\item \textbf{{Mobile device} location data:} The original {mobile device} phone location dataset is derived from SafeGraph~\cite{sg_social,sg_pattern}, a data company that aggregates anonymized GPS location data from numerous applications by census block group in order to provide insights about physical locations. The whole original dataset contains two major subdatasets: (i) social distancing metric and (ii) pattern of visits to Points of Interest (POIs). In the social distancing metric dataset, "home" is defined as the location of users at midnight, and "full-time workplace" is defined as the non-home location at which users spend more than $6$ hours during daytime. The pattern dataset mainly contains (i) base information of POIs of $168$ categories, including location name, address, brand association, etc., and (ii) information of daily visits and dwell time in POIs. 
	\\ { \textit{Purpose}: The rise in the number of COVID-19 cases indirectly influences electricity consumption through changes in individual behavior (social distancing), as well as policy-level changes (work-from-home or stay-at-home orders). We include mobile device location data to capture these influencing factors using two metrics - the {stay-at-home population} and the {population of on-site workers} (indicative of the social distancing level), and the {mobility in the retail sector}, indicative of the level of commercial activity. }
\end{itemize}
{Taken together, these cross-domain sources provide a more complete picture of the influencing factors and their complex multi-dimensional relationship that contributes to the changes in electricity consumption observed during COVID-19. }
\clearpage

\subsection*{{Supplementary Note SN-5: Remarks on Choice of Restricted VAR Model Parameters - Cases v.s. Hospitalizations/Deaths}}
{ 
	We first note the number of hospitalizations or deaths may be substituted for the number of COVID-19 cases without any change in the key findings (see Supplementary Table ST-4). However, we choose to calibrate the restricted VAR model with the number of confirmed cases (rather than deaths or hospitalizations) for the following reasons:
	\begin{itemize}
		\item First, the data on the number of COVID-19 cases is much easier to collect compared to other data such as the number of hospitalizations and ICU occupancy. The most comprehensive COVID-19 databases including John Hopkins University, the COVID Tracking Project, and the 1Point3Acres COVID Tracker only provide state-level hospitalization data. County-level data must be individually collected from the numerous county health department websites, and the availability of this data is not guaranteed. Further, among the seven hotspot cities that we focus on, only New York City and Chicago have such city-level data available. 
		\item Second, the number of deaths, while being more accurately reported than the number of COVID-19 cases, is a significantly delayed indicator; therefore, behavioral responses influencing electricity consumption may have already occurred in response to a rise in COVID-19 cases, before they are reflected in the number of deaths. Further, the number of deaths in some cities (Boston, Houston and Kansas City) are too small to be suitable for our statistical analysis (Supplementary Table ST-3).
		\item Third, under-reporting or inaccuracies in the reported number of COVID-19 cases are not of much concern in this model, since changes in electricity consumption are typically due to on individual and policy-level responses to the observed (reported) information. In fact, an analysis of the number of COVID-19 cases using Google Trends (Supplementary Fig. S-9) shows that was the number of COVID-19 cases was the most searched COVID-19 related keyword, occurring more than twice as often as COVID-19 deaths, indicating that the number of cases is a key factor in influencing the behavioral changes (social distancing and mobility), that result in changes in electricity consumption (Note that searches for COVID-19 hospitalizations and ICU occupancy were insignificant in comparison to the case numbers and deaths). 
		\item Last and most importantly, the modelling accuracy using the number of confirmed COVID-19 cases is sufficiently high from a statistical point of view. Further, we note that the core conclusions of the VAR model calibrated with the number of deaths, hospitalizations, or ICU occupancy data (Supplementary Tables ST-3 and ST-4) are essentially the same as those derived from the VAR model calibrated with COVID-19 case data. Moreover, using the number of deaths instead of the number of confirmed cases resulted in a lower fitting accuracy and potentially irregular impulse response directions (See RVAR-D in Supplementary Table ST-4). While the calibration of the VAR model with the number of hospitalizations resulted in higher accuracy, especially in Chicago, the model results are similar to those obtained by calibrating the VAR model using the number of confirmed cases.
	\end{itemize}
	Therefore, in summary,  we choose the number of confirmed cases as the metric to capture the severity of the pandemic due to its high availability, popularity and modelling performance.

}

\clearpage
\section*{Supplementary Methods}
The Statsmodels \cite{statsmodels} module in Python is used to implement the estimation of coefficients in the restricted Vector Autoregression (VAR) model and the corresponding statistical tests and analyses. We describe the key steps in estimating and verifying the restricted VAR model as follows.

\subsection*{Supplementary Method SM-1: Pre-estimation Preparation}
\subsubsection*{Data Pre-Processing}
Collecting electricity market data, weather data, COVID-19 public health data, {mobile device} location data, we pre-process the following variables as input candidates for the restricted VAR model:
\begin{itemize}
	\item \textbf{Load Reduction}: Logarithm of the amount of load reduction (MW) - samples with negative reduction (increase) are dropped for consistency.
	\item \textbf{New Daily Confirmed Cases}: Logarithm of the original count.
	\item \textbf{Stay-at-home Population}: Logarithm of the number of devices that stay at home completely.
	\item \textbf{Median Home Dwell Time Rate}: Median value of the home-dwelling time of sampled population.
	\item \textbf{Population of Full-time On-site Workers}: Logarithm of the net number of full-time on-site workers.
	\item \textbf{Population of Part-time On-site Workers}: Logarithm of the number of part-time on-site workers.
	\item \textbf{Mobility in the Retail Sector}: Logarithm of the number of visitors to retail locations as defined in Supplementary Note SN-3.
\end{itemize}

\subsubsection*{Augmented Dicket-Fuller Test}
The stationarity of time-series data is a prerequisite to calibrate a vector autoregression (VAR) model. Therefore, the Augmented Dickey-Fuller (ADF) test \cite{dickey1979distribution}, a commonly used unit root test, is used to test whether a time-series variable is non-stationary and possesses a unit root. The Augmented Dickey-Fuller test is carried out for each of the multiple time-series data that are candidate variables for restricted VAR model estimation. Each time-series is differenced to improve stationarity.



\subsubsection*{Cointegration Test}
The cointegration test\cite{engle1987coint} is used to test the long-term correlation between multiple non-stationary time-series. In addition to the ADF test which tests the stationarity of each de-trended input time-series, a cointegration test is used to detect potential long-term correlation among the original inputs, which is signified by the presence of cointergation. If cointegration is detected, the selected tuple of input time-series is not appropriate for restricted VAR modelling and needs to be dropped.

\subsubsection*{Granger Causality Wald Test}
Granger causality\cite{granger1969causal} is a probabilistic method to estimate the causality relationship among two variables represented as time-series. The key intuition behind the Granger Causality test is the assumption that future events cannot have causal effects on the past. To study the causality relationship between two variables, $x$ and $y$, we test if the lagged series $x_{t-n}, n \in Z^+$ will affect the current value $y_t$ for each time-step $t$, as concurrent and future values $x_{t+n}, n \in Z$ cannot affect $y_t$. The Granger Causality test is used to estimate the causal relationship between selected time-series variables shifted by the appropriate lag values. If significant counter-logical casual relationships are identified, we impose constraints on restricted VAR modeling such that the causalities are eliminated in the final restricted VAR model.
\clearpage

\subsection*{Supplementary Method SM-2: Restricted VAR Model Estimation}
After verifying the input variable time-series using the statistical tests described in Supplementary Method SM-1, we have a set of de-trended stationary input time-series that do not have long-term correlation among them. We model the dynamics of load reduction as a Vector Autoregression (VAR) model  of order $p$ as follows:
\begin{align}
	X_t=C+A_1X_{t-1}+\dots+A_pX_{t-p}+E_t,
\end{align}

where
\begin{align}
	A_i= \begin{bmatrix}
		a_{1,1}^i & a_{1,2}^i & \dots & a_{1,n}^i\\
		a_{2,1}^i & a_{2,2}^i & \dots & a_{2,n}^i\\
		\vdots & \vdots & \ddots  &  \vdots\\
		a_{n,1}^i & a_{n,2}^i & \dots & a_{n,n}^i
	\end{bmatrix},\quad
	X_t=\begin{bmatrix}
		x_t^1 \\
		x_t^2 \\
		\vdots \\
		x_t^n
	\end{bmatrix},\quad
	C=\begin{bmatrix}
		c^1 \\
		c^2 \\
		\vdots \\
		c^n
	\end{bmatrix},\quad
	E_t=\begin{bmatrix}
		e_t^1 \\
		e_t^2 \\
		\vdots \\
		e_t^n
	\end{bmatrix},
\end{align}
in which $C$ and $E_t$ are respectively column vectors of intercept and random errors, $x_t^1$ represents the target output variable at time $t$, namely the load reduction amount we wish to model, $x_t^2, ..., x_t^n$ represent the selected $n-1$ parameter variables including the number of COVID-19 cases, completely stay-at-home rate, median home dwell time rate, etc,  and the time notation $t-p$ represents the $p$-th lag of the variables. The coefficients $a_{i, j}^k$ are calculated separately for each variable using the Ordinary Least Square (OLS) estimator:
\begin{align}
	x_t^i = \sum_{j,k} a_{i,j}^k x^{t-k}_j+c_i.
\end{align}
Then, we can aggregate and concatenate all $a_{i, j}^k$ to obtain the regression matrix $A_k$, $0\leq k\leq p$.

If the result from the Granger causality test suggests that there may exist undesirable causal relationships between variables, we can impose constraints on the OLS to eliminate such relationships from appearing in the restricted VAR model. For example, if the Granger test suggests that variations in load reduction could be a causation of change in the stay-at-home population, which is intuitively illogical, we can restrict the corresponding entries of the VAR model coefficient matrices that describe such a relationship to zero during the OLS computation. With this Restricted VAR model, we ensure that the final model does not include unwanted relationships between parameter variables.

\clearpage

\subsection*{Supplementary Method SM-3: Restricted VAR Model Verification}
To verify the restricted VAR model, we need to guarantee stationarity, non-autocorrelation, and normality of the residual data, which is defined as
\begin{align}
	e_t = x_t-\sum_{i=1}^k\hat{A}_ix_{t-i}-c,
\end{align}
where $e_t$ and $x_t$ are the residuals and observation vectors respectively on day $t$, $k$ is the user-defined maximum lag, namely the order of the restricted VAR model, and $\{\hat{A}_i\}_{i=1}^k$ are the estimated coefficient matrices derived from the estimated restricted VAR model.

\subsubsection*{Augmented Dicket-Fuller Test for Residual Stationarity}
In addition to verifying the stationarity of the input time-series data in Supplementary Method SM-1, the ADF test is also used to check if the residual data are non-stationary and possess a unit root.

\subsubsection*{Ljung-Box Test for Residual Autocorrelation}
The endogeneity of the residual also needs to be verified, since the existence of endogeneity may render the regression result untrustworthy. Therefore, the Ljung-Box test \cite{box1970distribution,ljung1978measure } is used to test whether any of a group of autocorrelations of the residual time-series are different from zero. In this test, the null hypothesis is that the data are independently distributed while the alternative hypothesis is that the data are not independently distributed and exhibit serial correlation. The test statistic\cite{ljung1978measure} is defined as $Q=n(n+2)\sum_{i=1}^h\frac{\rho_i^2}{n-i}$, where $n$ is the number of residual data, $h$ is the number of lags selected to be $40$, and $\rho_i$ is the sample autocorrelation value at lag $i$. With significance level $5\%$, the critical region for rejection of the null hypothesis is $Q>\chi^2_{0.95,h}$.

\subsubsection*{Durbin-Watson Test for Residual Autocorrelation}
The Durbin-Watson statistic \cite{durbin1971testing} is another test statistic used to detect the presence of autocorrelation at lag $1$ in the residuals of the restricted VAR model. The null hypothesis is that the residuals are serially uncorrelated while the alternative is that they come from a first order autoregression process.


{\subsubsection*{Stability Test}
	The stability test is essentially a unit root test, by calculating the eigenvalues of the Restricted VAR model. We say that a Restricted VAR model is stable if the absolute values of all eigenvalues are equal to or less than $1$.}

\subsubsection*{Robustness Test for Parameter Stability}
The final step of examining a restricted VAR model is to test its robustness against parameter perturbations. The robustness of a given restricted VAR model is established by slightly changing each input parameter and perform the statistical tests such as AIC. We select the best model by considering the trade-off between robustness and accuracy and pick a model that has the best validation accuracy while demonstrating robustness in the test results.
\clearpage

\subsection*{Supplementary Method SM-4: Post-estimation Analysis}
With a restricted VAR model that passes all statistical tests, we need to analyze and interpret the results of the model. We perform the following analyses on the restricted VAR model. 
\subsubsection*{Impulse Response Analysis}
Impulse response analysis \cite{koop1996impulse} is an important step to describe the evolution of a restricted VAR model's variable in reaction to a shock in one or more variables. 

For the $p$-th order restricted VAR model of the form
\begin{align}
	x_t=c+A_1x_{t-1}+\dots+A_px_{t-p}+e_t,
\end{align}
where $x_t$, $c$ and $e_t$ are $n$ dimensional column vectors, $A_i$ is $p\times p$ dimensional matrix. In the impulse response analysis, we set the impulse response function $R(t)$ as 
\begin{align}
	R(t) = \sum_{i=1}^p A_i\cdot R(t-i),
\end{align}
with $R(0)=[0,\dots,1, \dots , 0]^T$ in which only one user-defined element equals to $1$.




It is particularly useful to consider the impulse response function as a predictive indicator that can forecast the dynamic behavior of electricity consumption in response to a change in any exogenous influencing factor, given a certain initial state. For example, suppose we have another model describing how public health policies impact social mobility, we can further combine these two models and simulate the impacts of public policies on the electricity consumption.

\subsubsection*{Forecast Error Variance Decomposition}
Forecast error variance decomposition (FEVD)\cite{lutkepohl2005new} is used to aid in the interpretation of the estimated restricted VAR model by determining the proportion of each variable's forecast error variance that is contributed by shocks to the other variables.

For the $p$-th order restricted VAR model in the form
\begin{align}
	x_t=c+A_1x_{t-1}+\dots+A_px_{t-p}+e_t,
\end{align}
where $x_t$, $c$ and $e_t$ are $n$ dimensional column vectors, $A_i$ is $p\times p$ dimensional matrix. Then, it can be reformulated as
\begin{align}
	X_t=C+AX_{t-1}+E_t,
\end{align}
where
\begin{align}
	X_t=\begin{bmatrix}
		x_t\\
		x_{t-1}\\
		x_{t-2}\\
		\vdots\\
		x_{t-p+1}
	\end{bmatrix},\quad
	C=\begin{bmatrix}
		c\\
		0\\
		0\\
		\vdots\\
		0
	\end{bmatrix},\quad
	A=\begin{bmatrix}
		A_1 & A_2 & \dots & A_{p-1} & A_p\\
		I_n    &     0 &  \dots &       0&     0\\
		0      &   I_n &  \dots &       0&     0\\
		\vdots &\vdots & \ddots & \vdots &\vdots\\
		0      &     0 &  \dots &    I_n &     0
	\end{bmatrix}\quad\text{and}\quad
	E_t=\begin{bmatrix}
		e_t\\
		0\\
		0\\
		\vdots\\
		0
	\end{bmatrix},
\end{align}
and $I_n$ is a $n\times n$ identity matrix, $A$ is a $np\times np$ matrix, and $X_t$, $C$ and $E_t$ are $np$ dimensional column vectors.

The mean square error of the $h$-step forecast of the $i$-th variable can be calculated as
\begin{align}
	\text{MSE}[x_{i}(h)]=[\sum_{j=0}^{h-1}\Phi_j\Sigma_e\Phi_j^T]_{(i,i)},
\end{align}
where $\Sigma_e$ is the covariance matrix of $e_t$, $\Phi_j=JA^jJ^T$, $J=[I_n,\quad 0,\quad \dots,\quad 0]_{k\times kp}$, $A^j$ is the $j$-th order power of $A$ matrix, and $(i,i)$ denotes the $i$-th diagonal element of the matrix.

Further, for the forecast error variance decomposition, $w_{ij}(h)$ is defined to represent the proportion of forecast error variance of the $i$-th variable at the $h$-th step accounted for by the shock to the $j$-th variable, that is, 
\begin{align}
	w_{ij}(h)=\frac{\sum_{k=0}^{h-1} (e_i^TB_ke_j)^2}{\text{MSE}[x_{i}(h)]},
\end{align}
where $e_i$ is the $i$-th columns of the identity matrix $I_{n\times n}$, $B_k=\Phi_kP$, and $P$ is a lower triangular matrix in the Cholesky factorization of $\Sigma_e$ such that $\Sigma_e=PP^T$. 

\clearpage

\subsection*{Supplementary Method SM-5: Restricted VAR Model Selection}
The aforementioned methods compute the coefficients of a restricted VAR model that is statistically robust. However, there is no explicit rule for selecting the parameter variables and the range of the training data. A numerical search on the parameter space is performed for each city to determine the optimal parameters. A list of parameters we use for iterative search is listed below.

\begin{itemize}
	\item Time-series from EMDA dataset used as input variables: $[x_1,...,x_n]$
	\item Date range of training data: $[T_1,T_2]$
	\item Order of the Restricted VAR model: $p$, ranging from $1$ to $7$
	\item Rule to determine whether to set a Restricted VAR coefficient to $0$: $r$
\end{itemize}

\noindent {Note that we have the following three rule candidates:}
\begin{enumerate}
	\item {Set $0$ all coefficients (except for the first row) in the first column of each matrix in the Restricted VAR model.}
	\item {In addition to the Rule 1, set the corresponding coefficients to $0$ if the Granger Causality Wald Test suggests that the $p$ value between two variables is over $0.1$.}
	\item {In addition to the Rule 1, set the corresponding coefficients to $0$ if the Granger Causality Wald Test suggests that the $p$ value between two variables is over $0.05$.}
\end{enumerate}

\noindent For each combination of possible parameters, the procedures described in Supplementary Methods SM-1,2,3,4 are performed to examine feasibility and quantify the numerical performance. After searching the parameter space, we determine the optimal combination of parameters according to the Akaike Information Criterion (AIC) \cite{sakamoto1986akaike} and Bayesian Information Criterion (BIC) \cite{chen1998speaker}. An ideal model should have low AIC and BIC value, and be able to to explain a large proportion of the variance in load reduction, while at the same time ensuring that the signs of the trends (increase/decrease) of all impulse responses are as desired.
The complete process of searching for the restricted VAR model is presented in Algorithm \ref{alg-1}:

\begin{algorithm}
	\caption{Iterative Search for VAR Parameters}
	\label{alg-1}
	\begin{algorithmic}
		\STATE Load complete training dataset into memory
		\STATE Compute all $n_p$ possible parameters combinations $\{[x_1,...,x_n],[T_1,T_2],p,[d_1,d_2],r\}$
		\FOR {parameter set $i=0$ to $n_p$}
		\STATE Difference the raw time-series array 
		\STATE Verify stationarity using ADF test
		\STATE Test for cointegration
		\STATE Perform Granger casuality test 
		\STATE Establish corresponding constraints for VAR computation
		\STATE Solve Restricted VAR using OLS and obtain coefficient matricies
		{
			\STATE Test for model stationarity and residual autocorrelation
			\STATE Quantify model performance using AIC and BIC information criteria}
		\ENDFOR
		\STATE Choose optimal parameter combinations and finalize model
	\end{algorithmic}
\end{algorithm}

\clearpage

\section*{{Supplementary Tables}}

\subsection*{{Supplementary Table ST-1: Restricted VAR Model Parameters}}
\begin{table}[htp!]
	\centering
	{   \begin{tabularx}{\textwidth}{XXXXX}
			\hline
			City & Start Date & End Date& Lags& Rule\\ \hline
			Boston & $03/25$ & $05/21$ & $4$ & $2$ \\
			Chicago & $04/06$ & $05/27$ & $4$ & $2$ \\
			Houston & $04/08$ & $06/28$ & $6$ & $2$ \\
			Kansas & $03/26$ & $06/03$ & $4$ & $2$ \\
			Los Angeles & $03/23$ & $05/23$ & $4$ & $2$ \\
			New York City & $03/24$ & $06/28$ & $6$ & $2$\\       Philadelphia& $03/24$ & $06/06$ & $6$ & $2$ \\\hline
	\end{tabularx}}
	\caption{{ Hyperparameters for the restricted VAR models of each city. Start and end date refer to the time range of the training data. Lags refer to the order of the restricted VAR models. Please see the definition of the Rule 2 in the Supplementary Method SM-5.}}
	\label{tab:test}
\end{table}
\clearpage

\subsection*{{Supplementary Table ST-2: VAR Statistical Test Results}}
\begin{table}[htp!]
	\centering
	{   \begin{tabularx}{\textwidth}{|X|X|X|X|X|X|X|}
			\hline
			City & Test & Load Reduction & COVID-19 daily confirmed cases& Stay-at-home population & On-site worker population & Retail mobility\\ \hline
			\multirow{5}{*}{Boston}
			& ADF test & $0.001$& $0.000$& $0.001$& $0.000$& $0.016$ \\\cline{2-7}
			& Cointegration & \multicolumn{5}{c|}{True}\\\cline{2-7}
			& LB test & $0.872$& $0.954$& $0.594$& $0.622$& $0.922$ \\\cline{2-7}
			& DW test & $2.073$& $1.956$& $1.940$& $2.220$& $1.899$ \\\cline{2-7} 
			& Stability & \multicolumn{5}{c|}{True}\\\hline\hline
			\multirow{5}{*}{Chicago}
			& ADF test & $0.000$& $0.000$& $0.000$& $0.000$& $0.000$ \\\cline{2-7}
			& Cointegration & \multicolumn{5}{c|}{True}\\\cline{2-7}
			& LB test & $0.879$& $0.991$& $0.609$& $0.472$& $0.756$ \\\cline{2-7}
			& DW test & $1.836$& $1.985$& $2.043$& $1.729$& $2.174$ \\\cline{2-7}
			& Stability & \multicolumn{5}{c|}{True}\\ \hline\hline
			\multirow{5}{*}{Houston}
			& ADF test & $0.000$& $0.000$& $0.027$& $0.001$& $0.001$ \\\cline{2-7}
			& Cointegration & \multicolumn{5}{c|}{True}\\\cline{2-7}
			& LB test & $0.863$& $0.997$& $0.873$& $0.895$& $0.738$ \\\cline{2-7}
			& DW test & $1.798$& $1.758$& $1.979$& $2.010$& $1.743$ \\ \cline{2-7}
			& Stability & \multicolumn{5}{c|}{True}\\\hline\hline
			\multirow{5}{*}{Kansas}
			& ADF test & $0.017$& $0.000$& $0.051$& $0.000$& $0.107$ \\\cline{2-7}
			& Cointegration & \multicolumn{5}{c|}{True}\\\cline{2-7}
			& LB test & $0.793$& $0.927$& $0.893$& $0.935$& $0.364$ \\\cline{2-7}
			& DW test & $2.065$& $2.117$& $2.053$& $2.200$& $2.326$ \\ \cline{2-7}
			& Stability & \multicolumn{5}{c|}{True}\\\hline\hline
			\multirow{5}{*}{Los Angeles}
			& ADF test & $0.016$& $0.000$& $0.051$& $0.000$& $0.107$ \\\cline{2-7}
			& Cointegration & \multicolumn{5}{c|}{True}\\\cline{2-7}
			& LB test & $0.934$& $0.966$& $0.978$& $0.703$& $0.834$ \\\cline{2-7}
			& DW test & $1.987$& $1.898$& $2.089$& $2.107$& $1.936$ \\ \cline{2-7}
			& Stability & \multicolumn{5}{c|}{True}\\\hline\hline
			\multirow{5}{*}{New York City}
			& ADF test & $0.000$& $0.026$& $0.000$& $0.000$& $0.001$ \\\cline{2-7}
			& Cointegration & \multicolumn{5}{c|}{True}\\\cline{2-7}
			& LB test & $0.932$& $0.979$& $0.867$& $0.973$& $0.873$ \\\cline{2-7}
			& DW test & $1.912$& $2.000$& $1.953$& $1.948$& $1.988$ \\ \cline{2-7}
			& Stability & \multicolumn{5}{c|}{True}\\\hline\hline
			\multirow{5}{*}{Philadelphia}
			& ADF test & $0.009$& $0.012$& $0.000$& $0.000$& $0.062$ \\\cline{2-7}
			& Cointegration & \multicolumn{5}{c|}{True}\\\cline{2-7}
			& LB test & $0.567$& $0.972$& $0.784$& $0.852$& $0.409$ \\\cline{2-7}
			& DW test & $1.808$& $2.115$& $2.044$& $2.133$& $2.056$ \\ \cline{2-7}
			& Stability & \multicolumn{5}{c|}{True}\\\hline
	\end{tabularx}}
	\caption{{ Statistical test results of the restricted VAR model of each city, including Augmented Dicket-Fuller (ADF), cointegration, Ljung-Box (LB), Durbin-Watson (DW), and stability tests. Note that this table shows the $p$ values of the ADF and LB tests, the statistics value of the DW test, and the True/False value of the cointegration and stability test. Here, the cointegration test result being True means that there is no variable having cointegration. The stability test result being True means that the Restricted VAR model is stable.}}
	\label{tab:test}
\end{table}

\clearpage

\subsection*{{Supplementary Table ST-3: Availability and Correlation Test of COVID-19 Public Health Metrics}}
\begin{table}[htp!]
	{ \begin{tabular}{cccccccc}
			\hline
			\textbf{City} & \textbf{Data}   & \textbf{Availability} & \textbf{Mean Value} & \multicolumn{3}{c}{\textbf{Quantiles}} & \textbf{Correlation Coefficient}  \\ 
			\cline{5-7}  &                 &                       &                     & 25\%       & 50\%        & 75\%        & \textbf{with Confirmed Case Data} \\ \hline
			New York City & Confirmed Case  & \textbf{\checkmark}            & 1789.58             & 384.75     & 1025.00     & 2892.50     & 1.000                             \\
			& Death           & \textbf{\checkmark}            & 191.80              & 24.75      & 63.00       & 304.25      & 0.902                             \\
			& Hospitalization & \textbf{\checkmark}            & 447.85              & 51.25      & 197.50      & 662.00      & 0.872                             \\ \hline
			Philadelphia  & Confirmed Case  & \textbf{\checkmark}            & 214.85              & 8.25       & 162.00      & 332.25      & 1.000                             \\
			& Death           & \textbf{\checkmark}            & 13.27               & 0.00       & 6.50        & 18.00       & 0.615                             \\
			& Hospitalization & \textbf{X}                    & ---                 & ---        & ---         & ---         & ---                               \\ \hline
			Boston        & Confirmed Case  & \textbf{\checkmark}            & 164.95              & 31.75      & 106.00      & 268.00      & 1.000                             \\
			& Death           & \textbf{\checkmark}            & 8.37                & 0.00       & 4.00        & 9.00        & 0.418                             \\
			& Hospitalization & \textbf{X}                    & ---                 & ---        & ---         & ---         & ---                               \\ \hline
			Chicago       & Confirmed Case  & \textbf{\checkmark}            & 961.40              & 507.50     & 915.50      & 1392.75     & 1.000                             \\
			& Death           & \textbf{\checkmark}            & 50.53               & 29.75      & 47.00       & 67.25       & 0.557                             \\
			& Hospitalization & \textbf{\checkmark}            & 1278.74             & 987.75     & 1400.50     & 1612.50     & 0.815                             \\ \hline
			Los Angeles   & Confirmed Case  & \textbf{\checkmark}            & 815.98              & 332.75     & 806.50      & 1255.00     & 1.000                             \\
			& Death           & \textbf{\checkmark}            & 27.54               & 7.50       & 25.00       & 45.00       & 0.551                             \\
			& Hospitalization & \textbf{X}                     & ---                 & ---        & ---         & ---         & ---                               \\ \hline
			Houston       & Confirmed Case  & \textbf{\checkmark}            & 243.97              & 56.00      & 170.00      & 289.00      & 1.000                             \\
			& Death           & \textbf{\checkmark}            & 3.09                & 0.00       & 2.00        & 5.00        & 0.494                             \\
			& Hospitalization & \textbf{X}                     & ---                 & ---        & ---         & ---         & ---                               \\ \hline
			Kansas City   & Confirmed Case  & \textbf{\checkmark}            & 18.26               & 2.00       & 12.00       & 27.00       & 1.000                             \\
			& Death           & \textbf{\checkmark}            & 0.67                & 0.00       & 0.00        & 1.00        & 0.097                             \\
			& Hospitalization & \textbf{X}                     & ---                 & ---        & ---         & ---         & ---                               \\ \hline
	\end{tabular}}
	\caption{{ Availability of different COVID-19 metrics in various hotspot cities and their correlation with COVID-19 confirmed case data.}}
\end{table}
\clearpage

\subsection*{{Supplementary Table ST-4: Statistical Tests of Restricted VAR Models Using Different COVID-19 Indicator Variables}}
\begin{table}[htp!]
	{ \begin{tabular}{cccccccc}
			\hline
			\textbf{City} & \textbf{Model} & \textbf{BIC}     & \textbf{AIC} & \textbf{Explainable Rate} & \textbf{Impulse Response-I} & \textbf{Impulse Response-II} \\ \hline
			New York City & RVAR-C         & -19.012          & -23.289      & 43.6\%       & Negative       & Positive \\
			& RVAR-D         & -18.729 & -23.122      & 44.1\%       & Positive       & Positive           \\
			& RVAR-H         & -19.499 & -23.776      & 36.1\%       & Negative       & Positive          \\ \hline
			Philadelphia  & RVAR-C         & -16.548 & -21.733      & 43.4\%       & Negative       & Positive      &           \\
			& RVAR-D         & -14.853 & -20.038      & 47.7\%       & Negative       & Positive          \\ \hline
			Chicago       & RVAR-C         & -16.978 & -20.675      & 26.1\%       & Negative       & Positive      &           \\
			& RVAR-D         & -16.217          & -19.914      & 24.4\%       & Negative       & Positive      &       \\
			& RVAR-H1        & -20.571 & -24.548      & 28.7\%       & Negative       & Positive          \\
			& RVAR-H2        & -20.435 & -24.132      & 15.6\%       & Negative       & Positive          \\ \hline
			Los Angeles   & RVAR-C         & -14.162          & -17.827      & 35.5\%       & Negative       & Positive         \\
			& RVAR-D         & -13.387 & -17.084      & 32.7\%       & Negative       & Positive   \\   \hline    
	\end{tabular}}\\
	
	\vspace{0.5em}
	{ RVAR-C: Restricted VAR model using the number of confirmed cases \\ RVAR-D: Restricted VAR model using the number of deaths  \\ RVAR-H / RVAR-H1: Restricted VAR model using the number of hospitalizations  \\ RVAR-H2: Restricted VAR model using the ICU occupancy numbers}
	\caption{{ Statistical results of restricted VAR models with different COVID-19 indicator variables. BIC and AIC are two information criteria that capture the model performance, as described in Supplementary SM-5; a model with a lower value of AIC and BIC performs better. the explainable rate shows how much of the variance in the electricity consumption can be explained by influencing factors other than its own trend. Impulse response-I is the electricity consumption change in response to COVID-19 related factors; a negative value means that the electricity consumption drops when this factor increases. Impulse response-II is the electricity consumption change in response to the mobility in the retail sector.}}
\end{table}

\clearpage
\bibliography{ref}
\end{document}